\documentclass[aps,preprint]{revtex4}%
\usepackage{amssymb}
\usepackage{amsfonts}
\usepackage{amsmath}
\usepackage{graphicx}%
\providecommand{\U}[1]{\protect\rule{.1in}{.1in}}
\newtheorem{theorem}{Theorem}

\newtheorem{definition}[theorem]{Definition}

\begin{document}
\title{Geometrical aspects of the Lie Algebra S-expansion Procedure}
\author{M. Artebani$^{2}$,\ R. Caroca$^{3}$, M. C. Ipinza$^{1}$, \ D. M.
Pe\~{n}afiel$^{1}$, P. Salgado$^{1}$}
\affiliation{$^{1}$Departamento de F\'{\i}sica, Universidad de Concepci\'{o}n Casilla
160-C, Concepci\'{o}n, Chile.}
\affiliation{$^{2}$Departamento de Matem\'{a}tica, Universidad de Concepci\'{o}n Casilla
\textbf{160-C}, Concepci\'{o}n, Chile.}
\affiliation{$^{3}$Departamento de Matem\'{a}tica y F\'{\i}sica Aplicadas, Universidad
Cat\'{o}lica de la Sant\'{\i}sima Concepci\'{o}n, Alonso de Rivera 285,
Concepci\'{o}n, Chile.}

\begin{abstract}
In this article it is shown that S-expansion procedure affects the geometry of
a Lie group, changing it and leading us to the geometry of another Lie group
with higher dimensionality. Is outlined, via an example, a method for
determining the semigroup, which would provide a Lie algebra from another.
Finally, it is proved that a Lie algebra obtained from another Lie algebra via
S-expansion is a non-simple Lie algebra.

\end{abstract}
\maketitle

\section{\textbf{Introduction}\bigskip}

In Ref. \cite{seagal} was pointed out that if two physical theories are
related by a limiting process, then the associated invariance groups should
also be related by some limiting process. This idea was studied in Ref.
\cite{iw} and introduced the so-called In\"{o}n\"{u}-Wigner contractions procedure.

Expansions of Lie algebras are a generalization of the contraction method and
were introduced some years ago in Refs. \cite{hs}, \cite{aipv}, \cite{irs},
\cite{irs1}. These procedure have been successfully applied in obtaining new
Lie algebras and the construction of gravitational theories \cite{gr-chs},
\cite{gr-bi}\textbf{. }

The procedure developed in references \cite{hs}, \cite{aipv} consists of
looking at the algebra $\mathcal{G}$ as described by the Maurer-Cartan forms
on the manifold of its associated group $G$ and, after rescaling some of the
group parameters by a factor $\lambda$, expanding the Maurer-Cartan forms as
series in $\lambda$. The expansion method, is different from the
In\"{o}n\"{u}-Wigner contraction method albeit, when the algebra dimension
does not change in the process, it may lead to a simple In\"{o}n\"{u}-Wigner
or In\"{o}n\"{u}-Wigner generalized contraction in the sense of Weimar-Woods
\cite{weim1}, \cite{weim2}.

On the other hand the method developed in references \cite{irs}, \cite{irs1}
is a natural outgrowth of the expansion method of Ref. \cite{aipv}. The
procedure is based on combining the structure constant of the algebra with the
inner law of a semigroup in order to define the Lie bracket of a new
$S$-expanded algebra. This Abelian Semigroup Expansion
method,\textquotedblleft S-expansion\textquotedblright, reproduces the results
of the Maurer-Cartan forms power series expansion for a particular choice of
the semigroup , but is formulated using the Lie algebra generators rather than
the associated Maurer-Cartan forms.

These methods appeared to be powerful tools in order to find non-trivial
relations between different Lie algebras. The discovery of these relations
presents in itself a very interesting problem from both physical and
mathematical points of view \cite{exp2}, \cite{gr-chs}, \cite{gr-bi}.

In this work it is shown that S-expansion procedure affects the geometry of a
Lie group. It is found how changing the magnitude of a vector and the angle
between two vectors. Is outlined, via an example, a method for determining the
semigroup, which would provide a Lie algebra from another. Finally, it is
proved that a Lie algebra obtained from another Lie algebra via S-expansion is
a non-simple Lie algebra.

The paper is organized as follows: In Sec. $II$ we review some concepts of the
theory of Lie algebras and \ \ \ \ \ the main aspects of the S-expansion
procedure. In Sec. $III$ we study \ \ how the S-expansion procedure affects
the geometry of a Lie group. It is found how changing the magnitude of a
vector and the angle between two vectors. \ In section $IV$ \ is outlined, via
an example, a method for determining the semigroup, which would provide a Lie
algebra from another\textbf{.}\ In section $V$ \ it is proved that a Lie
algebra obtained from another Lie algebra via S-expansion is a non-simple Lie
algebra. Here and in the following we have considered finite dimensional Lie
algebras and $K=%
\mathbb{C}
$ or $K=%
\mathbb{R}
$ as the fields involved.

\section{\textbf{Review some aspects of} \textbf{Lie Algebras and the
S-Expansion procedure}}

\subsection{\textbf{Some aspects of} \textbf{Lie Algebras}}

A Lie algebra is a linear vector space, but because of the group structure on
the manifold it inherits a rich algebraic structure. \ A Lie algebra
$\mathcal{G}$ is a vector space over a field $K$ on which a product $[,]$,
called the Lie bracket, is defined, with the properties%
\begin{equation}
\text{If }X,Y\in\mathcal{G}\text{, then }\left[  X,Y\right]  \in\mathcal{G}
\label{uno}%
\end{equation}

\begin{align}
\left[  \alpha X+\beta Y,Z\right]   &  =\alpha\left[  X,Z\right]
+\beta\left[  Y,Z\right]  \text{ for }\alpha,\beta\in K\text{ and }%
X,Y,Z\in\mathcal{G}.\label{dos.a}\\
\left[  X,\alpha Y+\beta Z\right]   &  =\alpha\left[  X,Y\right]
+\beta\left[  X,Z\right]  \text{ for }\alpha,\beta\in K\text{ and }%
X,Y,Z\in\mathcal{G}. \label{dos.b}%
\end{align}%
\begin{equation}
\left[  X,X\right]  =0\text{ \ \ for all }X\in\mathcal{G}. \label{tres}%
\end{equation}%
\begin{equation}
\left[  X,\left[  Y,Z\right]  \right]  +\left[  Y,\left[  Z,X\right]  \right]
+\left[  Z,\left[  X,Y\right]  \right]  =0. \label{cuatro}%
\end{equation}
The property (\ref{tres}) is called skew symmetry and (\ref{cuatro}) is know
as Jacobi identity.

If $\left\{  X_{i}\right\}  $ is a basis for $\mathcal{G}$ then we have%

\begin{equation}
\left[  X_{i},X_{j}\right]  =C_{ij}^{k}X_{k}, \label{7}%
\end{equation}
for some set of constants $C_{ij}^{k}$ called the structure constants of the
algebra. Accordingly, a Lie algebra may be specified by giving a set of
constants $C_{ij}^{k}$ such that%

\begin{equation}
C_{ij}^{k}=-C_{ji}^{k}\text{\ } \label{8}%
\end{equation}%
\begin{equation}
C_{ij}^{m}C_{mk}^{r}+C_{jk}^{m}C_{mi}^{r}+C_{ki}^{m}C_{mj}^{r}=0 \label{9'}%
\end{equation}

\textbf{Definition: }A representation of a Lie algebra $\mathcal{G}$ on a
vector space $V$ is a mapping $\rho$ from $\mathcal{G}$ to the linear
transformation of $V$ such that%
\begin{equation}
\rho\left(  \alpha X+\beta Y\right)  =\alpha\rho\left(  X\right)  +\beta
\rho\left(  Y\right)  \label{cinco}%
\end{equation}%
\begin{equation}
\rho\left(  \left[  X,Y\right]  \right)  =\left[  \rho\left(  X\right)
,\rho\left(  Y\right)  \right]  \label{sies}%
\end{equation}

\textbf{Transformation of basis: }Equations $\left(  1\right)  $ to $\left(
5\right)  $ do not uniquely determine the infinitesimal operators of a given
group. \ We are still free to replace the basis $Y_{i}$ by another. In fact,
under a change of basis transformation%

\begin{equation}
X_{i}=A_{i}^{\text{ }r}Y_{r} \label{9}%
\end{equation}
we find that the structure constants change as%

\begin{equation}
C_{rs}^{\prime\text{ \ }t}=\left(  A^{-1}\right)  _{r}^{\text{ }i}\left(
A^{-1}\right)  _{s}^{\text{ }j}C_{ij}^{k}A_{k}^{\text{ }t} \label{10}%
\end{equation}

Let $\mathcal{G}$ be a Lie algebra over the real numbers $%
\mathbb{R}
$ or the complex numbers $%
\mathbb{C}
$. Consider the linear map $adX$ of $\mathcal{G}$ into itself defined by
\begin{equation}
adX(Y)\equiv\left[  X,Y\right]  ,\text{ \ \ }X,Y\in\mathcal{G} \label{trece'}%
\end{equation}
Using the Jacobi identity (\ref{cuatro}), we get%
\begin{equation}
adX\left(  \left[  Y,Z\right]  \right)  =\left[  adX(Y),Z\right]  +\left[
Y,adX(Z)\right]  \label{trece}%
\end{equation}
i.e., the map $adX$ represents a derivation of $\mathcal{G}$. Furthemore,
using (\ref{trece'}) and the Jacobi identity we obtain%
\begin{equation}
ad\left[  X,Y\right]  (Z)=\left[  adX,adY\right]  \left(  Z\right)
\label{catorce}%
\end{equation}
Hence the set $\mathcal{G}_{a}=\left\{  adX,\text{ }X\in\mathcal{G}\right\}  $
is a linear Lie algebra, which is a subalgebra of the Lie algebra
$\mathcal{G}_{A}$ of all derivations and is called the $adjoint$ $algebra$.
The map $\psi:X\longrightarrow adX$ is the homomorphism of $\mathcal{G}$ onto
$\mathcal{G}_{a}$. \ 

It is easily verified that $\psi:X\longrightarrow adX$ is a representation of
the Lie algebra $\mathcal{G}$ with $\mathcal{G}$ itself considered as the
vector space of the representation. One need only check that $ad\left[
X,Y\right]  =\left[  adX,adY\right]  $, which is a simple consequence of the
Jacobi identity.

\subsubsection{\textbf{Adjoint representation}}

A better way to look at a change of basis transformation is to determine how
the change of basis affects the commutator of an arbitrary element $Z$ in the algebra%

\begin{equation}
\left[  Z,X_{i}\right]  =R(Z)_{i}^{\text{ }j}X_{j} \label{11}%
\end{equation}

Under the change of basis (\ref{9}) we find%

\[
\left[  Z,Y_{r}\right]  =S(Z)_{r}^{\text{ }s}Y_{s}%
\]%
\begin{align*}
\left[  Z,Y_{r}\right]   &  =\left[  Z,\left(  A^{-1}\right)  _{r}^{\text{ }%
i}X_{i}\right]  =\left(  A^{-1}\right)  _{r}^{\text{ }i}\left[  Z,X_{i}\right]
\\
S(Z)_{r}^{\text{ }s}Y_{s}  &  =\left(  A^{-1}\right)  _{r}^{\text{ }i}%
R(Z)_{i}^{\text{ }j}X_{j}=\left(  A^{-1}\right)  _{r}^{\text{ }i}%
R(Z)_{i}^{\text{ }j}A_{j}^{\text{ }s}Y_{s}%
\end{align*}
where
\begin{equation}
S(Z)_{r}^{\text{ }s}=\left(  A^{-1}\right)  _{r}^{\text{ }i}R(Z)_{i}^{\text{
}j}A_{j}^{\text{ }s}%
\end{equation}
In this manner the effect of a change of basis on the structure constants is
reduced to a study of similarity transformations.

The association of a matrix $R(Z)$ with each element of a Lie algebra is
called the regular or adjoint representation%

\begin{equation}
Z\overset{regular}{\underset{representation}{\longrightarrow}}R(Z). \label{13}%
\end{equation}
For example, the representation $adX$, called the adjoint representation,
always provides a matrix representation of the algebra. If $\left\{
X_{i}\right\}  $ is a basis for $\mathcal{G}$ then%
\[
adX_{i}(Y_{j})=\left[  X_{i},Y_{j}\right]  =R(X_{i})_{j}^{\text{ }k}%
X_{k}=C_{ij}^{k}X_{k}%
\]
Therefore the matrix associated with the transformation $adX_{i}$ is given by%
\[
R(X_{i})_{j}^{\text{ }k}=C_{ij}^{k}%
\]

\subsubsection{\textbf{Killing-Cartan inner product }}

The Killing-Cartan form of a Lie algebra is a symmetric bilinear form given by%

\begin{align}
\left(  X,X\right)   &  =tr\left(  R\left(  X\right)  R\left(  X\right)
\right)  =tr\left(  v^{i}R\left(  X_{i}\right)  v^{j}R\left(  X_{j}\right)
\right) \nonumber\\
&  =v^{i}v^{j}R\left(  X_{i}\right)  _{k}^{l}R\left(  X_{j}\right)  _{l}%
^{k}=v^{i}v^{j}\left(  C_{i}\right)  _{k}^{l}\left(  C_{j}\right)  _{l}^{k}.
\label{14}%
\end{align}

where $X=v^{i}X_{i}.$ This means that $v^{i}$ are coordinates in the algebra,
which fully define any arbitrary vector.

The inner product of Killing-Cartan provides information about the geometry of
the manifold of the group in a neighborhood of identity. The information is
obtained in terms of compactness, not compact or nilpotency of the group. This
information can be extrapolated to the rest of the manifold using the fact
that a Lie group is a "geodesically complete" manifold, i.e., we can
reconstruct it completely through the process of exponentiation of algebra.

The vector space of the Lie algebra can be divided into three subspaces under
the Cartan--Killing inner product. The inner product is positive-definite,
negative-definite, and identically zero. These three subspaces are denoted by:%

\[
\mathcal{G}=V_{-}+V_{+}+V_{0}%
\]

The subspace $V_{0}$ is a subalgebra of $\mathcal{G}$. It is the largest
nilpotent invariant subalgebra of $\mathcal{G}$. Under exponentiation, this
subspace maps onto the maximal nilpotent invariant subgroup in the original
Lie group.

The subspace $V_{-}$ is also a subalgebra of $\mathcal{G}$. It consists of
compact (a topological property) operators. That is to say, the exponential of
this subspace is a subset of the original Lie group that is parameterized by a
compact manifold. It also forms a subalgebra in $\mathcal{G}$ (not invariant).

Finally, the subspace $V_{+}$ is not a subalgebra of $\mathcal{G}$. It
consists of noncompact operators. The exponential of this subspace is
parameterized by a noncompact submanifold in the original Lie group.

The "division" of algebra in these subspaces implies that in the quadratic
form $\left(  X,X\right)  $ of the group, there are summands with different
signs representing the spaces denoted as $V_{-}$, $V_{+}$ and $V_{0}$.

\subsubsection{\textbf{Character of an algebra}}

The character of an algebra, denoted with the symbol $\chi$ measures the
degree of compactness of the manifold of the a group within a limited range of
integer values. The character of an algebra is defined as follows
\cite{gilmore}:%

\begin{equation}
\chi=\left(
\begin{array}
[c]{c}%
number\text{ }of\\
non\text{-}compact\\
generators
\end{array}
\right)  -\left(
\begin{array}
[c]{c}%
number\text{ }of\\
compact\\
generators
\end{array}
\right)  .
\end{equation}
which is the trace of the normalized Cartan--Killing form.

Consider a complex and semisimple Lie algebra (although it may contain sets of
nilpotent generators) $\mathcal{G}_{c}$ decomposed as follows%

\begin{equation}
X=\sum_{i=1}^{l}\mathcal{C}^{i}H_{i}+\sum_{\alpha\neq0}^{n-l}c^{\alpha
}E_{\alpha}%
\end{equation}
where $X\in\mathcal{G}_{%
\mathbb{C}
}$ and $\mathcal{C}$ is a complex coefficient. Here we have a decomposition of
$\mathcal{G}_{%
\mathbb{C}
}=\mathbf{H}\oplus\mathbf{E}$ type, where $\mathbf{H}$ is a compact subalgebra
$\mathcal{G}_{c}$ (maximal compact subalgebra). This expression is called
"Cartan decomposition" and basically split the algebra in two subspaces, one
with a negative definite Killing-Cartan metric and the other positive definite
(for semisimple case).

\subsection{\textbf{The S-expansion procedure}}

The expansion method proposed in Refs.\ \cite{hs}, \cite{aipv} consists in
considering the original algebra as described by its associated Maurer- Cartan
forms on the group manifold. Some of the group parameters are rescaled by a
factor $\lambda$ , and the Maurer-Cartan forms are expanded as a power series
in $\lambda$ . This series is finally truncated in a way that assures the
closure of the expanded algebra.

Consider now the main aspects of the $S$-expansion procedure and their
properties introduced in Ref. \cite{irs}. \ Let $S=\left\{  \lambda_{\alpha
}\right\}  $ be an abelian, discrete and finite semigroup with 2-selector
$K_{\alpha\beta}^{\ \ \ \gamma}$ defined by%
\begin{equation}
K_{\alpha\beta}^{\ \ \ \gamma}=\left\{
\begin{array}
[c]{cc}%
1 & \ \ \ \ \ \ \ \ \ \ \ \lambda_{\alpha}\lambda_{\beta}=\lambda_{\gamma}\\
0 & otherwise,
\end{array}
\right.
\end{equation}
and $\mathfrak{g}$ a Lie (super)algebra with basis $\left\{  \mathbf{T}%
_{A}\right\}  $ and structure constant $C_{AB}^{\ \ \ C}$,
\begin{equation}
\left[  \mathbf{T}_{A},\mathbf{T}_{B}\right]  =C_{AB}^{\ \ \ C}\mathbf{T}_{C}.
\end{equation}
Then it may be shown that the product $\mathfrak{G}=S\times\mathfrak{g}$ is
also a Lie (super)algebra with structure constants $C_{(A,\alpha)(B,\beta
)}^{\ \ \ \ \ \ \ \ \ \ \ \ (C,\gamma)}=K_{\alpha\beta}^{\ \ \gamma}%
C_{AB}^{\ \ \ \ C}$,
\begin{equation}
\left[  \mathbf{T}_{(A,\alpha)},\mathbf{T}_{(B,\beta)}\right]  =C_{(A,\alpha
)(B,\beta)}^{\ \ \ \ \ \ \ \ \ \ \ \ (C,\gamma)}\mathbf{T}_{(C,\gamma)}.
\end{equation}
The proof is direct and may be found in Ref. \cite{irs}.

\begin{definition}
Let $S$ be an abelian, discrete and finite semigroup and $\mathfrak{g}$ a Lie
algebra. The Lie algebra $\mathfrak{G}$ defined by $\mathfrak{G}%
=S\times\mathfrak{g}$ is called $S$-Expanded algebra of $\mathfrak{g}$.
\end{definition}

When the semigroup has a zero element $0_{S}\in S$, it plays a somewhat
peculiar role in the $S$-expanded algebra. The above considerations motivate
the following definition:

\begin{definition}
Let $S$ be an abelian semigroup with a zero element $0_{S}\in S$, and let
$\mathfrak{G}=S\times\mathfrak{g}$ be an $S$-expanded algebra. The algebra
obtained by imposing the condition $0_{S}\mathbf{T}_{A}=0$ on $\mathfrak{G}$
(or a subalgebra of it) is called $0_{S}$-reduced algebra of $\mathfrak{G}$
(or of the subalgebra).
\end{definition}

An $S$-expanded algebra has a fairly simple structure. Interestingly, there
are at least two ways of extracting smaller algebras from $S\times
\mathfrak{g}$. The first one gives rise to a \textit{resonant subalgebra},
while the second produces reduced algebras. \ In particular, a resonant
subalgebra can be obtained as follow.

Let $g=%
{\textstyle\bigoplus_{p\in I}}
V_{p}$ be a decomposition of $g$ in subspaces $V_{p}$, where $I$ is a set of
indices. \ For each $p,q\in I$ it is always possible to define $i_{\left(
p,q\right)  }\subset I$ such that%
\begin{equation}
\left[  V_{p},V_{q}\right]  \subset%
{\textstyle\bigoplus\limits_{r\in i_{\left(  p,q\right)  }}}
V_{r}, \label{eq33}%
\end{equation}
\textbf{\ }Now, let $S=%
{\textstyle\bigcup_{p\in I}}
S_{p}$ be a subset decomposition of the abelian semigroup $S$ such that%
\begin{equation}
S_{p}\cdot S_{q}\subset%
{\textstyle\bigcup_{r\in i_{\left(  p,q\right)  }}}
S_{p}. \label{eq34}%
\end{equation}
When such subset decomposition $S=%
{\textstyle\bigcup_{p\in I}}
S_{p}$ exists, then we say that this decomposition is in resonance with the
subspace decomposition of $g,$ $g=%
{\textstyle\bigoplus_{p\in I}}
V_{p}$.

The resonant subset decomposition is crucial in order to systematically
extract subalgebras from the $S$-expanded algebra $G=S\times g$, as is proven
in the following\medskip

\textbf{Theorem IV.2 of Ref.} \cite{irs}: Let $g=%
{\textstyle\bigoplus_{p\in I}}
V_{p}$ be a subspace decomposition of $g$, with a structure described by eq.
(\ref{eq33})$,$ and let $S=%
{\textstyle\bigcup_{p\in I}}
S_{p}$ be a resonant subset decomposition of the abelian semigroup $S$, with
the structure given in eq. (\ref{eq34}). Define the subspaces of $G=S\times
g$,%
\begin{equation}
W_{p}=S_{p}\times V_{p},\text{ \ }p\in I.
\end{equation}
Then,%
\begin{equation}
\mathfrak{G}_{R}=%
{\textstyle\bigoplus_{p\in I}}
W_{p}%
\end{equation}
is a subalgebra of $G=S\times g$.

Proof: \ the proof may be found in Ref. \cite{irs}.

\begin{definition}
The algebra $G_{R}=%
{\textstyle\bigoplus_{p\in I}}
W_{p}$ obtained is called a Resonant Subalgebra of the $S$-expanded algebra
$G=S\times g$.
\end{definition}

A useful property of the $S$-expansion procedure is that it provides us with
an invariant tensor for the $S$-expanded algebra $\mathfrak{G}=S\times
\mathfrak{g}$ in terms of an invariant tensor for $\mathfrak{g}$. As shown in
Ref. \cite{exp2} the theorem VII.2 provide a general expression for an
invariant tensor for a $0_{S}$-reduced algebra.

\textbf{Theorem VII.2 of Ref. \cite{irs}:} \ Let $S$ be an abelian semigroup
with nonzero elements $\lambda_{i}$, $i=0,\cdots,N$ and $\lambda_{N+1}=0_{S}$.
Let $\mathfrak{g}$ be a Lie (super)algebra of basis $\left\{  \mathbf{T}%
_{A}\right\}  $, and let $\langle\mathbf{T}_{A_{n}}\cdots\mathbf{T}_{A_{n}%
}\rangle$ be an invariant tensor for $\mathfrak{g}$. The expression
\begin{equation}
\langle\mathbf{T}_{(A_{1},i_{1})}\cdots\mathbf{T}_{(A_{n},i_{n})}%
\rangle=\alpha_{j}K_{i_{a}\cdots i_{n}}^{\ \ \ \ \ j}\langle\mathbf{T}_{A_{1}%
}\cdots\mathbf{T}_{A_{n}}\rangle
\end{equation}

where $\alpha_{j}$ are arbitrary constants, corresponds to an invariant tensor
for the $0_{S}$-reduced algebra obtained from $\mathfrak{G}=S\times
\mathfrak{g}$.

\textbf{Proof:} \ the proof may be found in section $VII$ of Ref. \cite{irs}.

In summary, in Refs. ~\cite{irs}, \cite{irs1}, \cite{exp2} was proposed a
natural outgrowth of power series expansion method, which is based on
combining the structure constant of the algebra ($\mathcal{G}$) with the inner
law of a semigroup $S$ in order to define the Lie bracket of a new
$S$-expanded algebra.

Theorem 1 of Ref. \cite{aipv} shows that, in the more general case, the
expanded Lie algebra has the structure constants%
\[
C_{(A,i)(Bj)}^{\text{ \ \ \ \ \ \ }(C,k)}=\left\{
\begin{array}
[c]{c}%
0\text{ \ \ when \ }i+j\neq k\\
C_{AB}^{\text{ \ \ \ \ }C}\text{ \ when \ }i+j=k
\end{array}
\right.
\]
where $i,j,k=0,\cdot\cdot\cdot,N$ \ correspond to the order of the expansion,
and $N$ is the truncation order. These structure constants can also be
obtained within the $S$-expansion procedure. In order to achieve this, one
must consider the $0_{S}$-reduction of an $S$-expanded algebra where $S$
corresponds to the semigroup. The Maurer-Cartan forms power series expansion
of an algebra $\mathcal{G}$, with truncation order $N$, coincides with the
$0_{S}$-reduction of the $S_{E}^{(N)}$-expanded algebra (see Ref. \cite{irs}).
\ \ This is of course no coincidence. The set of powers of the rescaling
parameter $\lambda$, together with the truncation at order $N$, satisfy
precisely the multiplication law of $S_{E}^{(N)}$. As a matter of fact, we
have $\lambda^{\alpha}\lambda^{\beta}=\lambda^{\alpha+\beta}$ and the
truncation can be imposed as $\lambda^{\alpha}=0$ when $\alpha>N$. \ It is for
this reason that one must demand $0_{S}T_{A}=0$ in order to obtain the
Maurer-Cartan expansion as an $S_{E}^{(N)}$-expansion: in this case the zero
of the semigroup is the zero of the field as well.

The $S$-expansion procedure is valid no matter what the structure of the
original $\mathcal{G}$ Lie algebra is, and in this sense it is very general.
However, when something about the structure of $\mathcal{G}$ is known, a lot
more can be done. As an example, in the context of Maurer Cartan expansion,
the rescaling and truncation can be performed in several ways depending on the
structure of $\mathcal{G}$, leading to several kinds of expanded algebras.
Important examples of this are the generalized In\"{o}n\"{u}--Wigner
contraction, or the $M$ algebra as an expansion of $osp(32|1)$ (see Refs.
\cite{aipv, deazcarraga2}). This is also the case in the context of
$S$-expansions. When some information about the structure of $\mathcal{G}$ is
available, it is possible to find subalgebras of $\mathcal{G}^{(E)}%
=S_{E}^{(N)}\times\mathcal{G}$ and other kinds of reduced algebras. Among
other examples we can find the obtention of General relativity from the
Maxwell algebras using the $0_{s}$-reduced-resonant procedure defined above
and showed in Refs. \cite{gr-chs, gr-bi, CPRS2, CPRS3}. In this way, all the
algebras obtained by the Maurer Cartan expansion procedure can be reobtained.
New kinds of $S$-expanded algebras can also be obtained by considering
semigroups different from $S_{E}^{(N)}$.

\section{\textbf{The expansion procedure and the geometry of a Lie group}}

\subsection{\textbf{Expanding the Killing-Cartan metric}}

The $S$-expansion procedure considers the product of an abelian, discrete and
finite semigroup $S=\left\{  \lambda_{1},\ldots,\lambda_{P}\right\}  $ and a
Lie algebra $\mathcal{G}$, which leads to a new Lie algebra generated by the
following $N\cdot P$ generators%

\begin{equation}
X_{\left(  \alpha,a\right)  }\equiv X_{A}=\left\{  X_{\left(  1,1\right)
},X_{\left(  1,2\right)  },\cdots,X_{\left(  1,N\right)  },\cdots,X_{\left(
P,N\right)  }\right\}
\end{equation}
where "$P"$ represents the number of elements of the semigroup and $N$
represents the number of generators of the Lie algebra $\mathcal{G}.$

Now consider the metric of the new space and its intrinsic geometric
properties. If $X$ is a vector of the vector space $S\otimes\mathcal{G}$, then
we introduce the inner product as the Killing-Cartan product (for details see
Appendix $A$)
\begin{equation}
\left(  X,X\right)  _{S-\exp}\equiv tr\left(  R\left(  X\right)  R\left(
X\right)  \right)  .
\end{equation}

Therefore, we have%

\begin{align}
\left(  X,X\right)  _{S-\exp}\equiv tr\left(  R\left(  X\right)  R\left(
X\right)  \right)   &  =tr\left(  v^{\left(  \alpha,a\right)  }R\left(
X_{\left(  \alpha,a\right)  }\right)  v^{\left(  \beta,b\right)  }R\left(
X\right)  _{\left(  \beta,b\right)  }\right) \nonumber\\
&  =v^{\left(  \alpha,a\right)  }v^{\left(  \beta,b\right)  }R\left(
X_{\left(  \alpha,a\right)  }\right)  _{\left(  \gamma,c\right)  }^{\left(
\delta,d\right)  }R\left(  X_{\left(  \beta,B\right)  }\right)  _{\left(
\delta,d\right)  }^{\left(  \gamma,c\right)  }\nonumber\\
&  =v^{\left(  \alpha,a\right)  }v^{\left(  \beta,b\right)  }K_{\alpha\gamma
}^{\delta}\left(  C_{a}\right)  _{c}^{d}K_{\beta\delta}^{\gamma}\left(
C_{b}\right)  _{d}^{c}\nonumber\\
&  =v^{\left(  \alpha,a\right)  }v^{\left(  \beta,b\right)  }K_{\alpha\gamma
}^{\delta}K_{\beta\delta}^{\gamma}~tr\left(  R\left(  X_{a}\right)  R\left(
X_{b}\right)  \right) \nonumber\\
&  =v^{\left(  \alpha,a\right)  }v^{\left(  \beta,b\right)  }K_{\alpha\gamma
}^{\delta}K_{\beta\delta}^{\gamma}~\left(  X_{a},X_{b}\right)
\end{align}

Hence we see that the product of Killing-Cartan, undergoes a change due to the
presence of the $K$-selectors.

Because the Killing-Cartan product, of the original algebra appears immersed
in the Killing-Cartan product of the expanded algebra, the calculations are
simplified. Since a metric is a symmetric bilinear form, we can use the
spectral theorem, to obtain the corresponding diagonal metric. (This theorem
states that every real symmetric matrix is diagonalizable in $%
\mathbb{R}
$).

This means that a transformation $X\longrightarrow\tilde{X}$ allows to write
$\left(  \tilde{X},\tilde{X}\right)  =0,$\ $\forall$\ $a\neq b$. So that%

\begin{equation}
\left(  \tilde{X},\tilde{X}\right)  =\tilde{v}^{a}\tilde{v}^{a}\left(
\tilde{X}_{a},\tilde{X}_{a}\right)  ,\nonumber
\end{equation}

therefore the Killing-Cartan product of the expanded algebra, takes the form%

\begin{equation}
\left(  \tilde{X},\tilde{X}\right)  _{S-\exp}=\tilde{v}^{\left(
\alpha,a\right)  }\tilde{v}^{\left(  \beta,a\right)  }K_{\alpha\gamma}%
^{\delta}K_{\beta\delta}^{\gamma}~\left(  \tilde{X}_{a},\tilde{X}_{a}\right)
~\text{.}%
\end{equation}

This means%

\begin{equation}
\left(  \tilde{X},\tilde{X}\right)  _{S-\exp}=\tilde{v}^{\left(
\alpha,a\right)  }\tilde{v}^{\left(  \beta,b\right)  }K_{\alpha\gamma}%
^{\delta}K_{\beta\delta}^{\gamma}~\left(  \tilde{X}_{a},\tilde{X}_{b}\right)
\neq0\text{, \ \ when }a=b\text{.}%
\end{equation}

This new inner product is invariant under the action of the S-expanded
transformations (S-expanded generators) that constituing the new Lie algebra
$S\otimes\mathcal{G}$ (A proof of invariance is given in Appendix $B$). This
is because the original inner product $\left(  X_{a},X_{b}\right)  $ is
invariant under the $\mathcal{G}$ tranformations and the part that involve the
$K$ selectors doesn%
\'{}%
t affect the inner product $v^{\left(  \alpha,a\right)  }v^{\left(
\beta,b\right)  }K_{\alpha\gamma}^{\delta}K_{\beta\delta}^{\gamma}~\left(
X_{a},X_{b}\right)  $ because of the abelian and the associativity properties
of the semigroup $S$.

However not all matrices are diagonalizable. The decomposition of a matrix in
Jordan canonical form is a decomposition that generalizes the notion of
diagonalization. The interesting thing about this decomposition is that every
matrix can be taken to its canonical form, i.e., any matrix $A$ can be written
in the form%

\[
A=SJS^{-1},
\]
where $J$, is known as Jordan matrix.

This means that for each value of the index "$a$", the values that can take
the pair of indices $\left\{  \alpha,\beta\right\}  $ will play the role of
labeling new coordinates unlike the index set $\left\{  \gamma,\delta\right\}
$ whose role is of geometrical nature. For this reason we use the indices
$\left\{  i,j\right\}  $ to denote $\left\{  \alpha,\beta\right\}  $. Thus, we have%

\begin{align}
\left(  \tilde{X},\tilde{X}\right)  _{S-\exp}=  &  \sum_{a}^{\dim\left(
\mathcal{G}\right)  }~\left[  \sum_{i,j,\gamma,\delta}^{p}\tilde{v}^{\left(
i,a\right)  }\tilde{v}^{\left(  j,a\right)  }K_{i\gamma}^{\delta}K_{j\delta
}^{\gamma}\right]  ~\left(  \tilde{X}_{a},\tilde{X}_{a}\right) \nonumber\\
\left(  \tilde{X},\tilde{X}\right)  _{S-\exp}  &  =\sum_{a}^{\dim\left(
\mathcal{G}\right)  }~\left[  \sum_{i,j}^{p}\tilde{v}^{\left(  i,a\right)
}\tilde{v}^{\left(  j,a\right)  }\left[  \sum_{\gamma,\delta}^{p}K_{i\gamma
}^{\delta}K_{j\delta}^{\gamma}\right]  \right]  ~\left(  \tilde{X}_{a}%
,\tilde{X}_{a}\right)  .
\end{align}

If of all the bases that diagonalize the original metric, we choose the basis
of eigenvectors, we have%

\begin{align}
\left(  \tilde{X},\tilde{X}\right)  _{S-\exp}  &  =\sum_{a}^{\dim\left(
\mathcal{G}\right)  }~\left[  \sum_{i,j}^{P}\tilde{v}^{\left(  i,a\right)
}\tilde{v}^{\left(  j,a\right)  }\left[  \sum_{\gamma,\delta}^{P}K_{i\gamma
}^{\delta}K_{j\delta}^{\gamma}\right]  \right]  ~\left(  \tilde{X}_{a}%
,\tilde{X}_{a}\right) \\
&  \longrightarrow\sum_{a}^{\dim\left(  \mathcal{G}\right)  }~\left[
\sum_{i,j}^{P}\hat{v}^{\left(  i,a\right)  }\hat{v}^{\left(  j,a\right)
}\left[  \sum_{\gamma,\delta}^{P}K_{i\gamma}^{\delta}K_{j\delta}^{\gamma
}\right]  \right]  ~\lambda_{a}%
\end{align}
where the change $\tilde{v}\longrightarrow\hat{v}$ means the rotation of the
base under $O\left(  n\right)  $ and $\lambda_{a}$ corresponds to the
eigenvalue associated to the eigenvector $X_{\lambda_{a}}$. Note that although
the original metric is diagonal, the expanded metric need not be diagonal.
This is because the coordinates of the new $S\otimes\mathcal{G}$ vector space,
denoted as $\tilde{v}^{\left(  \alpha,a\right)  }$, has indices of the
semigroup $S$. Thus we have%

\begin{align}
\left(  X,X\right)  _{S-\exp}  &  =\left(  \tilde{v}^{\left(  \alpha,a\right)
}\right)  _{\substack{1 \\\times\left(  P\cdot N\right)  }}%
\begin{pmatrix}
\lambda_{1}\left(  \sum_{\gamma,\delta}^{p}K_{\alpha\gamma}^{\delta}%
K_{\beta\delta}^{\gamma}\right)  &  & O\\
& \ddots & \\
O &  & \lambda_{N}\left(  \sum_{\gamma,\delta}^{p}K_{\alpha\gamma}^{\delta
}K_{\beta\delta}^{\gamma}\right)
\end{pmatrix}
_{\substack{\left(  P\cdot N\right)  \\\times\left(  P\cdot N\right)
}}\left(  \tilde{v}^{\left(  \beta,b\right)  }\right)  _{\substack{\left(
P\cdot N\right)  \\\times1}}\nonumber\\
&  \equiv\left(  \tilde{v}^{\left(  \alpha,a\right)  }\right)  _{\substack{1
\\\times\left(  P\cdot N\right)  }}%
\begin{pmatrix}
\lambda_{1}\left(  M_{K}\right)  _{p\times p} &  & O\\
& \ddots & \\
O &  & \lambda_{N}\left(  M_{K}\right)  _{p\times p}%
\end{pmatrix}
_{\substack{\left(  P\cdot N\right)  \\\times\left(  P\cdot N\right)
}}\left(  \tilde{v}^{\left(  \beta,b\right)  }\right)  _{\substack{\left(
P\cdot N\right)  \\\times1}}\nonumber\\
&  \equiv\left(  \tilde{v}^{\left(  \alpha,a\right)  }\right)  _{1\times
\left(  P\cdot N\right)  }~~\left(  g\right)  _{\left(  \alpha,a\right)
\left(  \beta,b\right)  }~~\left(  \tilde{v}^{\left(  \beta,b\right)
}\right)  _{\left(  P\cdot N\right)  \times1}%
\end{align}

From here, we can see that the intrinsic local geometry of the variety of the
expanded group depends strongly on the characteristics of the semigroup.

The matrix $\left(  M_{K}\right)  _{p\times p}$ defined as $\left(
M_{K}\right)  _{p\times p}=K_{\alpha\gamma}^{\delta}K_{\beta\delta}^{\gamma},$
is symmetric because the semigroup is abelian. This allows writing%

\begin{equation}
\left(  X,X\right)  _{S-\exp}=\left(  \tilde{v}^{\left(  \alpha,a\right)
}\right)  _{1\times\left(  P\cdot N\right)  }%
\begin{pmatrix}
\lambda_{1}\left(  M_{K}\right)  _{p\times p} &  & O\\
& \ddots & \\
O &  & \lambda_{N}\left(  M_{K}\right)  _{P\times P}%
\end{pmatrix}
_{\left(  P\cdot N\right)  \times\left(  P\cdot N\right)  }\left(  \tilde
{v}^{\left(  \beta,a\right)  }\right)  _{\left(  P\cdot N\right)  \times
1}\nonumber
\end{equation}

or,%
\[
\left(  X,X\right)  _{S-\exp}=
\]

\[
\left(  \tilde{v}^{\left(  \alpha,a\right)  }\right)  _{\substack{1
\\\times\left(  P\cdot N\right)  }}%
\begin{pmatrix}
V.P.
\end{pmatrix}
_{\substack{\left(  P\cdot N\right)  \\\times\left(  P\cdot N\right)  }}%
\begin{pmatrix}
\lambda_{1}\bar{\lambda}_{1} &  &  &  &  &  & \\
& \ddots &  &  &  & O & \\
&  & \lambda_{1}\bar{\lambda}_{p} &  &  &  & \\
&  &  & \ddots &  &  & \\
&  &  &  & \lambda_{N}\bar{\lambda}_{1} &  & \\
& O &  &  &  & \ddots & \\
&  &  &  &  &  & \lambda_{N}\bar{\lambda}_{P}%
\end{pmatrix}
_{\substack{\left(  P\cdot N\right)  \\\times\left(  P\cdot N\right)  }}%
\begin{pmatrix}
V.P.
\end{pmatrix}
_{\substack{\left(  P\cdot N\right)  \\\times\left(  P\cdot N\right)  }%
}^{-1}\left(  \tilde{v}^{\left(  \beta,a\right)  }\right)  _{\substack{\left(
P\cdot N\right)  \\\times1}}
\]

where%

\begin{equation}%
\begin{pmatrix}
V.P.
\end{pmatrix}
_{\left(  p\cdot N\right)  \times\left(  p\cdot N\right)  }=%
\begin{pmatrix}
V.P.\left(  \lambda_{1}\right)  & \cdots & V.P.\left(  \lambda_{N\cdot
p}\right)
\end{pmatrix}
\end{equation}

is the matrix of eigenvectors associated with the eigenvalues $\bar{\lambda}$
of a symmetric matrix, denoted as $\left(  g\right)  _{AB}$.

This result is very important in, for example, the study of change in the
signature of the metric with respect to the signature of the original metric.
Since the entries of these submatrices are formed by elements of the form
$K_{\alpha\gamma}^{\delta}K_{\beta\delta}^{\gamma}$ will always be possible to
modify them using the laws of the internal composition of the semigroup $S$.

To see how it affected the metric of signature $\left(  +_{1},+_{2}%
,\cdots,+_{l},\cdots,-_{1},-_{2},\cdots,-_{m}\right)  $ with $l+m=N$, rewrite
$\left(  g_{AB}\right)  $ as%

\[
_{d}\left(  g_{AB}\right)  =
\]

\begin{equation}%
\begin{pmatrix}
+\lambda_{1}\left(  \pm\bar{\lambda}_{1}\right)  &  &  &  &  &  &  &  &  & \\
& \ddots &  &  &  &  &  &  &  & \\
&  & +\lambda_{1}\left(  \pm\bar{\lambda}_{p}\right)  &  &  &  &  & O &  & \\
&  &  & \ddots &  &  &  &  &  & \\
&  &  &  & +\lambda_{l}\left(  \pm\bar{\lambda}_{p}\right)  &  &  &  &  & \\
&  &  &  &  & -\lambda_{l+1}\left(  \pm\bar{\lambda}_{1}\right)  &  &  &  & \\
&  &  &  &  &  & \ddots &  &  & \\
&  & O &  &  &  &  & -\lambda_{l+1}\left(  \pm\bar{\lambda}_{p}\right)  &  &
\\
&  &  &  &  &  &  &  & \ddots & \\
&  &  &  &  &  &  &  &  & -\lambda_{l+m}\left(  \pm\bar{\lambda}_{p}\right)
\end{pmatrix}
_{\substack{N\cdot p~\times\\N\cdot p}}
\end{equation}
so if the matrix $\left(  M_{K}\right)  _{p\times p}$ has for example a
negative eigenvalue $\pm\bar{\lambda}_{p}\longrightarrow-\bar{\lambda}_{p}$,
it is found%

\begin{equation}
_{d}\left(  g_{AB}\right)  =%
\begin{pmatrix}
+\lambda_{1}\left(  \pm\bar{\lambda}_{1}\right)  &  &  &  &  &  &  &  &  & \\
& \ddots &  &  &  &  &  &  &  & \\
&  & -\lambda_{1}\bar{\lambda}_{p} &  &  &  &  & O &  & \\
&  &  & \ddots &  &  &  &  &  & \\
&  &  &  & -\lambda_{l}\bar{\lambda}_{p} &  &  &  &  & \\
&  &  &  &  & -\lambda_{l+1}\left(  \pm\bar{\lambda}_{1}\right)  &  &  &  & \\
&  &  &  &  &  & \ddots &  &  & \\
&  & O &  &  &  &  & +\lambda_{l+1}\bar{\lambda}_{p} &  & \\
&  &  &  &  &  &  &  & \ddots & \\
&  &  &  &  &  &  &  &  & +\lambda_{l+m}\bar{\lambda}_{p}%
\end{pmatrix}
_{\substack{N\cdot p \\\times N\cdot p}}
\end{equation}

So there will be a change of sign for each of the $N$ matrices matrices
$\left(  M_{K}\right)  _{p\times p}$. This means that the dimension of the
original algebra, as well as the dimension of its subspaces plays a crucial
role in the study of the final signature of the metric. If the dimension of
the metric is $N=2n$ for some $n\in%
\mathbb{N}
$ and $p\in\left\{
\mathbb{N}
\right\}  $, then necessarily there will be an even number of sign changes in
the diagonal: Even numbers form a semigroup under multiplication and similarly
for $N=2n+1.$

A "change of sign" means that the signs have changed due to the presence of
negative eigenvalues in $M_{K}$ matrix. That is, are changes with respect to
the case where the signature of the diagonal depends only on the eigenvalues
of the original metric. Denote the number of sign changes in the metric
S-expanded with the symbol $\#_{-}$. We will use the following notation for
algebra $\mathcal{G}$ of $\dim\left(  \mathcal{G}\right)  =N$%

\begin{align}
number~of~matrices\text{ }\left(  M_{K}\right)  _{p\times p}  &  =N\nonumber\\
number~of~eigenvalues\text{ }of~g_{ab}  &  =N\nonumber\\
number\text{ }of\text{ }elements\text{ }of\text{ }the\text{ }semigroup  &
=P\nonumber\\
number\text{ }of\text{ }diagonal\text{ }elements\text{ }of~_{d}\left(
X,X\right)  _{S-\exp}  &  =N\cdot P\\
number\text{ }of\text{ }negative\text{ }eigenvalues\text{ }of~\left(
M_{K}\right)  _{p\times p}  &  =Q\leq P\nonumber\\
Total\text{ }number\text{ }of\text{ }changes\text{ }of\text{ }sign  &  =N\cdot
Q\text{.}\nonumber
\end{align}

In the number of diagonal elements of $_{d}\left(  X,X\right)  _{S-\exp
}=N\cdot P$ should consider the zeros that come from the original algebra,
which may have nilpotent subalgebras or subspaces.

Consider now the distribution of number $\#_{-}$ along the diagonal
$_{d}\left(  X,X\right)  _{S-\exp}$. To see this it is necessary first to
clarify the increase in the dimensionality of each subspace of the vector
space of the original $\mathcal{G}$ algebra, which will be differentiated by
the corresponding diagonal signature. They will be denoted by $V_{+}$
subspaces whose diagonal is totally positive and $V_{-}$ the subspaces whose
diagonal is totally negative. There are also spaces (nilpotent) $V_{0}$ whose
metric has only zeros.

So we have to add to the above table the quantities%

\begin{align}
ran\left(  V_{+}\right)   &  =l\nonumber\\
ran\left(  V_{-}\right)   &  =m\nonumber\\
ran\left(  V_{+}\right)  _{S-\exp}  &  =l\cdot P\\
ran\left(  V_{-}\right)  _{S-\exp}  &  =m\cdot P\nonumber
\end{align}

This allows us to make a detailed analysis of each of the subspaces of
$\mathcal{G}$. Indeed, it is straightforward to see that%

\begin{align*}
change\text{ }of\text{ }sign\text{ }in~\left(  V_{+}\right)  _{s-\exp}  &
=l\cdot Q\\
change\text{ }of\text{ }sign\text{ }in~~\left(  V_{-}\right)  _{s-\exp}  &
=m\cdot Q\\
change\text{ }of\text{ }sign\text{ }in~~\left(  \mathcal{G}\right)  _{s-\exp}
&  =\left(  l+m\right)  \cdot Q\\
&  =N\cdot Q
\end{align*}

Note that, if \ $ran\left(  V_{+}\right)  =ran\left(  V_{-}\right)  $, \ then%

\[
\#_{-}\left(  V_{+}\right)  _{S-\exp}=l\cdot Q=m\cdot Q=\#_{-}\left(
V_{-}\right)  _{S-\exp}%
\]

and that if $ran\left(  V_{+}\right)  \neq ran\left(  V_{-}\right)  $, \ then%

\[
\#_{-}\left(  V_{+}\right)  _{S-\exp}=l\cdot Q\neq m\cdot Q=\#_{-}\left(
V_{-}\right)  _{S-\exp}%
\]

as well as if $ran\left(  V_{+}\right)  >ran\left(  V_{-}\right)
~~$or$~~ran\left(  V_{+}\right)  <ran\left(  V_{-}\right)  ,$ then%

\begin{align*}
\#_{-}\left(  V_{+}\right)  _{S-\exp}  &  =l\cdot Q~>~m\cdot Q=\#_{-}\left(
V_{-}\right)  _{S-\exp}~~~\\
or~~~\#_{-}\left(  V_{+}\right)  _{S-\exp}  &  =l\cdot Q~<~~m\cdot
Q=\#_{-}\left(  V_{-}\right)  _{S-\exp}%
\end{align*}

Now consider a classification of algebras, based on the above results%

\begin{align}
I  &  =\left\{  ran\left(  \mathcal{G}\right)  =2n~~con~n\in%
\mathbb{N}
\text{ }:ran\left(  V_{+}\right)  =ran\left(  V_{-}\right)  \right\}
\nonumber\\
II  &  =\left\{  ran\left(  \mathcal{G}\right)  =2n~~con~n\in%
\mathbb{N}
\text{ }:ran\left(  V_{+}\right)  \neq ran\left(  V_{-}\right)  \right\} \\
III  &  =\left\{  ran\left(  \mathcal{G}\right)  =2n+1~~con~n\in%
\mathbb{N}
\right\}  \text{ .}\nonumber
\end{align}

In the study of changes of sign in the expanded metrics, $ran\left(
\mathcal{G}\right)  $ plays an interesting role. In fact, consider the
analysis of each of these three sets when the $M_{K}$ matrix has $Q$ negative eigenvalues:

\paragraph{Case $I$ with $\dim\left(  V_{0}\right)  =0:$}

In this case,%

\begin{equation}%
\begin{array}
[c]{ccc}%
ran\left(  V_{+}\right)  =l=ran\left(  V_{-}\right)  =m, & where &
l~and~m~are~even\\
ran\left(  V_{+}\right)  _{S-\exp}=l\cdot P=ran\left(  V_{-}\right)  _{S-\exp
}=m\cdot P & where & l\cdot P~and~m\cdot P~are~even
\end{array}
\end{equation}

\begin{align}
ran\left(  \mathcal{G}\right)   &  =l+m:\text{ }even\nonumber\\
ran\left(  \mathcal{G}\right)  _{S-\exp}  &  =\left(  l+m\right)  \cdot
P=ran\left(  \mathcal{G}\right)  \cdot P:even,\text{ }\forall\text{ \ }P\in%
\mathbb{N}%
\end{align}

This means that these algebras conserve parity under $S$-expansion because
even numbers have the property of closure under addition and multiplication.
In presence of negative eigenvalues producing sign changes it is found%

\[
_{d}\left(  g_{AB}\right)  _{I}=%
\begin{pmatrix}
\Gamma & O\\
O & \Lambda
\end{pmatrix}
\]
where%

\begin{align}
_{d}\left(  g_{AB}\right)  _{\Gamma}  &  =%
\begin{pmatrix}
\lambda_{1}{}_{d}\left(  M_{K}\right)  &  &  &  &  & \\
& \ddots &  &  & O & \\
&  & \ddots &  &  & \\
&  &  & \lambda_{i~d}\left(  M_{K}\right)  &  & \\
& O &  &  & \ddots & \\
&  &  &  &  & \lambda_{l=m}~_{d}\left(  M_{K}\right)
\end{pmatrix}
_{\substack{\left(  l\cdot P\right)  \times\left(  l\cdot P\right)  \\=
\\\left(  m\cdot P\right)  \times\left(  m\cdot P\right)  }}\nonumber\\
_{d}\left(  g_{AB}\right)  _{\Lambda}  &  =%
\begin{pmatrix}
-\lambda_{l+1~d}\left(  M_{K}\right)  &  &  &  & \\
& \ddots &  & O & \\
&  & -\lambda_{l+i}~_{d}\left(  M_{K}\right)  &  & \\
& O &  & \ddots & \\
&  &  &  & -\lambda_{l+m}~_{d}\left(  M_{K}\right)
\end{pmatrix}
_{\substack{\left(  m\cdot P\right)  \times\left(  m\cdot P\right)  \\=
\\\left(  l\cdot P\right)  \times\left(  l\cdot P\right)  }}\nonumber
\end{align}

where%

\begin{equation}
_{d}\left(  M_{K}\right)  =%
\begin{pmatrix}
-\bar{\lambda}_{1} &  &  &  &  & \\
& \ddots &  &  & O & \\
&  & -\bar{\lambda}_{Q} &  &  & \\
&  &  & \bar{\lambda}_{Q+1} &  & \\
& O &  &  & \ddots & \\
&  &  &  &  & \bar{\lambda}_{P}%
\end{pmatrix}
\end{equation}

In this case it is found that for each of the $l$ positive eigenvalues of
$_{d}\left(  X,X\right)  $, are produced $Q$ changes of sign of the form
$+\longrightarrow-$. This means that are produced in total $l\cdot
Q\equiv\left(  \#_{-}\right)  _{\Gamma}$ changes. The same is true for the
$\Lambda$ submatrix but with $m\cdot Q\equiv\left(  \#_{-}\right)  _{\Lambda}$
changes the type $-\longrightarrow+$. So we have%

\begin{align}
ran\left(  V_{+}\right)  \cdot Q+ran\left(  V_{-}\right)  \cdot Q  &
=2\frac{ran\left(  \mathcal{G}\right)  }{2}\cdot Q\\
&  =ran\left(  \mathcal{G}\right)  \cdot Q=2n\cdot Q=\#_{-}%
\end{align}
changes of sign.

The important thing here is how this change is distributed in $_{d}\left(
X,X\right)  _{S-\exp}$. Algebras set to $I$, the amount of negative elements
equals the number of positive elements. So the difference between these
amounts (difference between the number of positive and negative diagonal
elements without regard to the numerical value) is given by $\chi=\lambda
_{+}-\lambda_{-}=0=\chi_{S-\exp}$.

It should be noted that although $\#_{-}$ is an even number important thing is
how the change of sign is distributed in the diagonal elements of $_{d}\left(
X,X\right)  _{S-\exp}$. This distribution generates changes in the number
$\chi$ and plays an important role in the general classification of real forms
of a complex Lie algebra.

\subsubsection{\textbf{Case of the sets II and III}}

In these cases,%

\[
_{d}\left(  g_{AB}\right)  _{I}=%
\begin{pmatrix}
\Gamma & O\\
O & \Lambda
\end{pmatrix}
\]

where%

\begin{align}
_{d}\left(  g_{AB}\right)  _{\Gamma}  &  =%
\begin{pmatrix}
\lambda_{1d}\left(  M_{K}\right)  &  &  &  &  & \\
& \ddots &  &  & O & \\
&  & \lambda_{id}\left(  M_{K}\right)  &  &  & \\
&  &  & \lambda_{i+1}\left(  \bar{\lambda}_{Q+1}\right)  &  & \\
& O &  &  & \ddots & \\
&  &  &  &  & \lambda_{l\neq m}\left(  \bar{\lambda}_{P}\right)
\end{pmatrix}
_{\substack{\left(  l\cdot P\right)  \times\left(  l\cdot P\right)
\\\neq\\\left(  m\cdot P\right)  \times\left(  m\cdot P\right)  }}\nonumber\\
_{d}\left(  g_{AB}\right)  _{\Lambda}  &  =%
\begin{pmatrix}
-\lambda_{l+1~~d}\left(  M_{K}\right)  &  &  &  &  & \\
& \ddots &  &  & O & \\
&  & -\lambda_{l+i~~d}\left(  M_{K}\right)  &  &  & \\
&  &  & -\lambda_{l+i+1}~_{d}\left(  \bar{\lambda}_{Q+1}\right)  &  & \\
& O &  &  & \ddots & \\
&  &  &  &  & -\lambda_{l+m~~d}\left(  M_{K}\right)
\end{pmatrix}
_{\substack{\left(  m\cdot P\right)  \times\left(  m\cdot P\right)
\\\neq\\\left(  l\cdot P\right)  \times\left(  l\cdot P\right)  }}\nonumber
\end{align}

where%

\[
_{d}\left(  M_{K}\right)  =%
\begin{pmatrix}
-\bar{\lambda}_{1} &  &  &  &  & \\
& \ddots &  &  & O & \\
&  & -\bar{\lambda}_{Q} &  &  & \\
&  &  & \bar{\lambda}_{Q+1} &  & \\
& O &  &  & \ddots & \\
&  &  &  &  & \bar{\lambda}_{P}%
\end{pmatrix}
\]
in this case occur $l\cdot Q=\left(  \#_{-}\right)  _{\Gamma}$ changes of sign
$+\longrightarrow-$, and $m\cdot Q=\left(  \#_{-}\right)  _{\Lambda}$ changes
of sign \ $-\longrightarrow+$, in a total of%

\begin{align*}
\#_{\Gamma}+\#_{\Lambda}  &  =ran\left(  V_{+}\right)  \cdot Q+ran\left(
V_{-}\right)  \cdot Q\\
&  =l\cdot Q+m\cdot Q\\
&  =ran\left(  \mathcal{G}\right)  \cdot Q=2n\cdot Q=\#_{-}%
\end{align*}
where we can see that the value $\#$ is similar to the case of the set $I$,
for the same value of $n\in%
\mathbb{N}
$. However the outcome of interest in this case is%

\begin{align}
l\cdot Q  &  \neq m\cdot Q\nonumber\\
&  \Longrightarrow\nonumber\\
\left(  \#_{-}\right)  _{\Gamma}  &  \neq\left(  \#_{-}\right)  _{\Lambda
}\nonumber\\
&  \Longrightarrow\nonumber\\
\chi_{I}  &  =\left(  \lambda_{+}-\lambda_{-}\right)  _{I}\neq\left(
\lambda_{+}-\lambda_{-}\right)  _{II}=\chi_{II}\text{ .}%
\end{align}
This is due to the fact that the transformation in the signature of
$_{d}\left(  X,X\right)  _{S-\exp}$ appears differently in each subspace, when
the original algebra is of type $II$ and $III$.

Consider now the relationship between the elements of the set $\left\{
ran\left(  V_{+}\right)  ,~ran\left(  V_{-}\right)  ,~P,~Q\right\}  $ during
the process of change in the value of $\chi$ difference. To find this
relationship let's see what happens during the process:

\begin{itemize}
\item Increase of rank of $V_{+}$%
\[
ran\left(  V_{+}\right)  \longrightarrow ran\left(  V_{+}\right)  \cdot P
\]

\item Reduction in the rank under the change of internal sign%
\[
ran\left(  V_{+}\right)  \cdot P\longrightarrow ran\left(  V_{+}\right)  \cdot
P-\left(  \#_{-}\right)  _{\Gamma}%
\]

\item Increase of the rank due to internal change signature of another
subspace
\[
ran\left(  V_{+}\right)  \cdot P-\left(  \#_{-}\right)  _{\Gamma
}\longrightarrow ran\left(  V_{+}\right)  \cdot P-\left(  \#_{-}\right)
_{\Gamma}+\left(  \#_{-}\right)  _{\Lambda},
\]

\end{itemize}

that rewritten in terms of the elements of the set $\left\{  ran\left(
V_{+}\right)  ,~ran\left(  V_{-}\right)  ,~P,~Q\right\}  $ we have
(considering semisimplicity)%

\begin{align*}
ran\left(  V_{+}\right)  \cdot P-\left(  \#_{-}\right)  _{\Gamma}+\left(
\#_{-}\right)  _{\Lambda}  &  =ran\left(  V_{+}\right)  \cdot P-ran\left(
V_{+}\right)  \cdot Q+ran\left(  V_{-}\right)  \cdot Q\\
&  =ran\left(  V_{+}\right)  \cdot P-\chi\cdot Q
\end{align*}

So the character of algebra $\chi_{S-\exp}$ is given by%

\begin{align}
\chi_{S-\exp}  &  =ran\left(  V_{+}\right)  _{S-\exp}-ran\left(  V_{-}\right)
_{S-\exp}\nonumber\\
&  =ran\left(  V_{+}\right)  \cdot P-ran\left(  V_{+}\right)  \cdot
Q+ran\left(  V_{-}\right)  \cdot Q\nonumber\\
&  -\left(  ran\left(  V_{-}\right)  \cdot P-ran\left(  V_{-}\right)  \cdot
Q+ran\left(  V_{+}\right)  \cdot Q\right) \nonumber\\
&  =\left(  ran\left(  V_{+}\right)  -ran\left(  V_{-}\right)  \right)
\cdot\left(  P-2Q\right) \nonumber\\
&  =\chi\cdot\left(  P-2Q\right)  \text{.}%
\end{align}

For nilpotent subspaces if they exist we have the following additional process

\begin{itemize}
\item Increase in the dimensionality of $V_{0}$%
\[
V_{0}\longrightarrow V_{0}\cdot P
\]

\item Increase in the dimensionality because the occurrence of zeros in
$V_{\pm}$.
\begin{equation}
V_{0}\longrightarrow V_{0}\cdot p\longrightarrow V_{0}\cdot P+\left(
V_{+}+V_{-}\right)  \cdot H
\end{equation}

\end{itemize}

where "$H$" denotes the increase.

From this we can see that for Lie algebras of the set $I$ have%

\begin{equation}
\chi_{S-\exp}=\chi\cdot\left(  P-2Q\right)  =0\nonumber
\end{equation}
that is, regardless of the order semigroup or the amount of eigenvalues
negative or zero of the matrix $M_{K}$ the value of the difference
$\chi_{S-\exp}$ will always be zero. So, we have that algebras of type I will
keep type I.

For the sets $II$, $III$ is necessary to consider the following points

\begin{itemize}
\item The set of natural numbers (including zero) forms a group under addition
but it is not group under the operation of subtraction: That is, it is not
necessarily true that%
\[
\left\{  a,b\right\}  \in%
\mathbb{N}
^{\ast}\text{ }/~a-b\in%
\mathbb{N}
^{\ast}%
\]

\item The set of even numbers form a group under addition and subtraction.

\item The set of odd numbers do not form a group under subtraction nor under addition.

\item An even number plus an odd number is always odd.
\end{itemize}

This means that the sets $II$ and $III$ satisfy the following properties:%

\begin{align}
\underset{\in~\left\{  2%
\mathbb{Z}
+1\right\}  }{\underbrace{ran\left(  V_{+}\right)  }}+\underset{\in~\left\{  2%
\mathbb{Z}
+1\right\}  }{\underbrace{ran\left(  V-\right)  }}~  &  \in II\nonumber\\
\underset{\in~\left\{  2%
\mathbb{Z}
\right\}  }{\underbrace{ran\left(  V_{+}\right)  }}+\underset{\in~\left\{  2%
\mathbb{Z}
\right\}  }{\underbrace{ran\left(  V-\right)  }}~  &  \in II
\end{align}

\begin{align}
\underset{\in~\left\{  2%
\mathbb{Z}
\right\}  }{\underbrace{ran\left(  V_{+}\right)  }}+\underset{\in~\left\{  2%
\mathbb{Z}
+1\right\}  }{\underbrace{ran\left(  V-\right)  }}~  &  \in III\nonumber\\
\underset{\in~\left\{  2%
\mathbb{Z}
+1\right\}  }{\underbrace{ran\left(  V_{+}\right)  }}+\underset{\in~\left\{  2%
\mathbb{Z}
\right\}  }{\underbrace{ran\left(  V-\right)  }}~  &  \in III
\end{align}

and then, independently of the semigroup, we have%

\begin{align*}
\chi_{S-\exp}  &  =\chi\cdot\left(  P-2Q\right)  \underset{I}{\longrightarrow
}\chi_{S-\exp}=0\cdot\left(  P-2Q\right) \\
\chi_{S-\exp}  &  =\chi\cdot\left(  P-2Q\right)  \underset{II}{\longrightarrow
}\chi_{S-\exp}=\left(  2m-2n\right)  \cdot\left(  P-2Q\right) \\
\chi_{S-\exp}  &  =\chi\cdot\left(  P-2Q\right)  \underset{III}%
{\longrightarrow}\chi_{S-\exp}=%
\genfrac{\{}{.}{0pt}{0}{\left[  2m+1-2n\right]  \cdot\left(  P-2Q\right)
}{\left[  2m-2n-1\right]  \cdot\left(  P-2Q\right)  }%
\\
con~~m,n  &  \in%
\mathbb{N}%
\end{align*}
\bigskip when $ran\left(  V_{\pm}\right)  \neq0$ .

This shows that $\chi_{S-\exp}$ depends both the original $\chi$ value and
semigroup characteristics.

An interesting question is: when it will produce a $\chi_{S-\exp}$ with
different sign to the original $\chi$ ?. To answer consider the following results:

\begin{itemize}
\item If%
\[
\left(  \mp\right)  \chi_{S-\exp}=\left(  \pm\right)  \chi\cdot\left(
P-2Q\right)  ,
\]

\end{itemize}

that is%

\begin{equation}
P-2Q<0,
\end{equation}

then%

\[
\frac{P}{2}<Q~.
\]

$2)$ If%

\[
\left(  \pm\right)  \chi_{S-\exp}=\left(  \pm\right)  \chi\cdot\left(
P-2Q\right)  ,
\]

that is%

\begin{equation}
P-2Q>0
\end{equation}

then%

\[
\frac{P}{2}>Q~.
\]

In other words, when $ran\left(  V_{\pm}\right)  \neq0$ and the matrix has a
number $Q$ of negative eigenvalues greater than half of the quantity $P$
(semigroup elements), a change will occur in the signature of the type
$\left(  \pm\right)  \chi\longrightarrow\left(  \mp\right)  \chi_{S-\exp}$
independently of the original algebra.

Furthermore, when the $M_{K}$ matrix has a $Q$ number of negative eigenvalues
smaller than half of the of $P$ (we most remember that the condition $P/2\in%
\mathbb{N}
$ has to be satisfied.), then there will be a process of the type $\left(
\pm\right)  \chi\longrightarrow\left(  \pm\right)  \chi_{S-\exp}$
independently of the original algebra.

Consider now the study of the intrinsic geometry for the case that $M_{K}$ has
null eigenvalues $\bar{\lambda}=0$.

Consider the following process

\begin{itemize}
\item Increasing the $V_{+}$ dimensionality of, for example,
\[
ran\left(  V_{+}\right)  \longrightarrow ran\left(  V_{+}\right)  \cdot P
\]

\item Decreased dimensionality due to the appearance of internal zeros%
\[
ran\left(  V_{+}\right)  \cdot P\longrightarrow ran\left(  V_{+}\right)  \cdot
P-ran\left(  V_{+}\right)  \cdot H=ran\left(  V_{+}\right)  \cdot\left(
P-H\right)
\]

\item The dimensionality of $V_{+}$, is not affected by the occurrence of zero
elements in the complementary space, in this case $V_{-}$%
\[
ran\left(  V_{+}\right)  \cdot P-ran\left(  V_{+}\right)  \cdot
H\longrightarrow ran\left(  V_{+}\right)  \cdot P-ran\left(  V_{+}\right)
\cdot H
\]

\end{itemize}

thus $\chi_{S-\exp}$ in the case $\bar{\lambda}=0,$ is given by%

\begin{align}
\chi_{S-\exp}  &  =ran\left(  V_{+}\right)  _{S-\exp}-ran\left(  V_{-}\right)
_{S-\exp}\nonumber\\
&  =ran\left(  V_{+}\right)  \cdot P-ran\left(  V_{+}\right)  \cdot H-\left[
ran\left(  V_{-}\right)  \cdot P-ran\left(  V_{-}\right)  \cdot H\right]
\nonumber\\
&  =\chi\cdot\left(  P-H\right)
\end{align}

for $H\leq P$ always.

In this case it is not possible a process of the type $\left(  \pm\right)
\chi\longrightarrow\left(  \mp\right)  \chi_{S-\exp}$ because the condition
$H\leq P$, impedes that $\chi\cdot\left(  P-H\right)  $ having a different
sign from the sign of $\chi$.

\bigskip

We should also note that if a matrix has eigenvalues null then a decrease in
their range occurs. This can be seen from the fact that%

\begin{equation}
ran\left(  \mathcal{G}\right)  _{S-\exp}=d\cdot P
\end{equation}
where "$d~$" denotes the number of non-zero diagonal elements of $_{d}\left(
X,X\right)  $. The amount ($ran\left(  \mathcal{G}\right)  _{S-\exp} $)
decrease if the rank of the matrix $M_{K}$ decreases. Since there are
$N=\dim\left(  \mathcal{G}\right)  $ matrices $M_{K}$ in $\left(  X,X\right)
_{S-\exp}$, we have that for each eigenvalue zero of $M_{K}$ matrices, there
will be a decrease in the total range given by%

\begin{align}
decrease\text{ }of\text{ }the\text{ }rank  &  =H\cdot N\\
with~H  &  \leq P\nonumber
\end{align}

\begin{align*}
ran\left(  _{d}\left(  X,X\right)  _{S-\exp}\right)   &  =ran\left(
_{d}\left(  X,X\right)  \right)  ~\cdot P-_{d}\left(  X,X\right)  \cdot H\\
&  =~ran\left(  _{d}\left(  X,X\right)  \right)  \cdot\left(  P-H\right)  ~.
\end{align*}

For $\dim\left(  V_{0}\right)  =0,$ we have%

\[
ran\left(  \mathcal{G}\right)  _{S-\exp}=ran\left(  \mathcal{G}\right)
\left(  P-H\right)
\]
which allows to calculate the expression for the change in the amount
$\chi_{S-\exp}$ in the presence of positive, negative and zero eigenvalues
($\dim\left(  V_{\pm}\right)  \neq0$)%

\begin{align}
\chi_{S-\exp}  &  =ran\left(  V_{+}\right)  \cdot P-ran\left(  V_{+}\right)
\cdot Q-ran\left(  V_{+}\right)  \cdot H+ran\left(  V_{-}\right)  \cdot
Q\nonumber\\
&  -\left[  ran\left(  V_{-}\right)  \cdot P-ran\left(  V_{-}\right)  \cdot
Q-ran\left(  V_{-}\right)  \cdot H+ran\left(  V_{+}\right)  \cdot Q\right]
\nonumber\\
&  =\chi\left(  P-H-2Q\right)  \text{.}%
\end{align}

Let $n_{+},s_{+},N_{+}$ denote the number of positive eigenvalues,
$n_{-},s_{-},N_{-}$ the number of negative eigenvalues and $n_{0},s_{0},N_{0}$
the number of zero eigenvalues (all eigenvalues are counted with
multiplicity). That is, with respect to the notation used: $P=s_{+}%
+s_{-}+s_{0}$; $Q=s_{-}$ and $H=s_{0}$ ;\ Moreover $\chi_{S-\exp}:=s_{+}%
-s_{-}$. \ The discussion about the signature of the S-expansion given above,
could be resumed by the following theorem, in terms of the signatures of the
original algebra and of the matrix $M$ associated to the semigroup. \ 

\ 

\textbf{Theorem: }Let $\mathcal{G}$ be a real Lie algebra of dimension $n$
whose Killing form has signature $\left(  n_{+},n_{-},n_{0}\right)  $ and let
$S$ a semigroup of order $s$ whose associated matrix $M_{k}$ has signature
$\left(  s_{+},s_{-},s_{0}\right)  $. Then the Killing form of the
$S-$expanded algebra $\mathcal{G}_{S-\exp}$ is $\left(  N_{+},N_{-}%
,N_{0}\right)  $, where%

\begin{equation}
N_{+}=n_{+}s_{+}+n_{-}s_{-}~~,~\text{ }N_{-}=n_{-}s_{+}+n_{+}s_{-}%
~~,~~N_{0}=ns_{0}+sn_{0}~.
\end{equation}

In particular, the rank of the form for the $S-$expansion is%

\begin{equation}
rank\left(  S\times\mathcal{G}\right)  =N_{+}+N_{-}=\left(  n_{+}%
+n_{-}\right)  \left(  s_{+}+s_{-}\right)  =rank\left(  M_{k}\right)
rank\left(  \mathcal{G}\right)
\end{equation}

and%

\[
\chi_{\left(  S\times\mathcal{G}\right)  }=N_{+}-N_{-}=\left(  n_{+}%
-n_{-}\right)  \left(  s_{+}-s_{-}\right)  =\chi_{_{M_{k}}}~\chi_{\mathcal{G}}%
\]

\textit{Proof. }This follows from the diagonalized form of the matrix for the
Killing form of the $S-$expanded Lie algebra giving at the begining of this
section for the matrix associated to de equation $\left(  35\right)  $, wich
shows that the eigenvalues are of the form $\lambda_{i}\bar{\lambda}_{j}$,
where $\lambda_{i}$ is an eigenvalue for $\mathcal{G}$ and $\bar{\lambda}_{j}$
an eigenvalue for $M_{k}$. This implies that a positive eigenvalue is either
the product of two positive or two negative eigenvalues. A negative eigenvalue
is the product of a negative and a positive eigenvalue, or viceversa. Finally
the zero eigenvalue is obtained as the product of a zero eigenvalue with any
other eigenvalue.

\ 

\textbf{Example: }Consider the following simple example. Using $\chi=-1$,
$ran\left(  V_{+}\right)  =1$, $ran\left(  V_{-}\right)  =2$, $P=4$, $Q=3$, we
find that the expanded metric takes the form%

\[
_{d}\left(  X,X\right)  _{S-\exp}=%
\begin{pmatrix}
D_{+} &  & O\\
& D_{-} & \\
O &  & D_{-}%
\end{pmatrix}
\]

where%

\[
D_{\pm}=\pm~\lambda~\left(  \pm\bar{\lambda}\right)  \cdot\left(  I\right)
_{ran\left(  V_{\pm}\right)  \cdot P~\times~ran\left(  V_{\pm}\right)  \cdot
P}~.
\]

Since $P=4$ and $Q=3$, we can write%

\begin{align*}
D_{+}  &  =%
\begin{pmatrix}
+\lambda\left(  +\bar{\lambda}\right)  &  &  & O\\
& +\lambda\left(  -\bar{\lambda}\right)  &  & \\
&  & +\lambda\left(  -\bar{\lambda}\right)  & \\
O &  &  & +\lambda\left(  -\bar{\lambda}\right)
\end{pmatrix}
=%
\begin{pmatrix}
+\lambda\left(  +\bar{\lambda}\right)  &  &  & O\\
& -\lambda\bar{\lambda} &  & \\
&  & -\lambda\bar{\lambda} & \\
O &  &  & -\lambda\bar{\lambda}%
\end{pmatrix}
\\
D_{-}  &  =%
\begin{pmatrix}
-\lambda\left(  +\bar{\lambda}\right)  &  &  & O\\
& -\lambda\left(  -\bar{\lambda}\right)  &  & \\
&  & -\lambda\left(  -\bar{\lambda}\right)  & \\
O &  &  & -\lambda\left(  -\bar{\lambda}\right)
\end{pmatrix}
=%
\begin{pmatrix}
-\lambda\left(  +\bar{\lambda}\right)  &  &  & O\\
& +\lambda\bar{\lambda} &  & \\
&  & +\lambda\bar{\lambda} & \\
O &  &  & +\lambda\bar{\lambda}%
\end{pmatrix}
\end{align*}

which leads to the following change in the dimensionality of the subspaces%

\begin{align*}
ran\left(  V_{+}\right)  _{S-\exp}  &  =1\cdot P-1\cdot Q+2\cdot
Q=1\cdot4-1\cdot3+2\cdot3=7\\
ran\left(  V_{-}\right)  _{S-\exp}  &  =2\cdot P-2\cdot Q+1\cdot
Q=2\cdot4-2\cdot3+1\cdot3=5
\end{align*}
so that%

\[
\chi_{S-\exp}=ran\left(  V_{+}\right)  _{S-\exp}-ran\left(  V_{-}\right)
_{S-\exp}=7-5=2~.
\]

Thus%

\[
\chi_{S-\exp}=\chi\cdot\left(  P-2Q\right)  =-1\cdot\left(  4-6\right)
=-1\cdot-2=2~.
\]

In the case that $P=4$, $Q=2$, $H=1$, we \ have%

\begin{align*}
D_{+}  &  =%
\begin{pmatrix}
+\lambda\left(  +\bar{\lambda}\right)  &  &  & O\\
& +\lambda\left(  -\bar{\lambda}\right)  &  & \\
&  & +\lambda\left(  -\bar{\lambda}\right)  & \\
O &  &  & +\lambda\left(  0\right)
\end{pmatrix}
=%
\begin{pmatrix}
+\lambda\left(  +\bar{\lambda}\right)  &  &  & O\\
& -\lambda\bar{\lambda} &  & \\
&  & -\lambda\bar{\lambda} & \\
O &  &  & 0
\end{pmatrix}
\\
D_{-}  &  =%
\begin{pmatrix}
-\lambda\left(  +\bar{\lambda}\right)  &  &  & O\\
& -\lambda\left(  -\bar{\lambda}\right)  &  & \\
&  & -\lambda\left(  -\bar{\lambda}\right)  & \\
O &  &  & -\lambda\left(  0\right)
\end{pmatrix}
=%
\begin{pmatrix}
-\lambda\left(  +\bar{\lambda}\right)  &  &  & O\\
& +\lambda\bar{\lambda} &  & \\
&  & +\lambda\bar{\lambda} & \\
O &  &  & 0
\end{pmatrix}
\end{align*}
from where%
\begin{align*}
ran\left(  V_{+}\right)  _{S-\exp}  &  =ran\left(  V_{+}\right)  \cdot
P-ran\left(  V_{+}\right)  \cdot H-ran\left(  V_{+}\right)  \cdot Q+ran\left(
V_{-}\right)  \cdot Q\\
&  =1\cdot4-1\cdot1-1\cdot2+2\cdot2=5\\
ran\left(  V_{-}\right)  _{S-\exp}  &  =ran\left(  V_{-}\right)  \cdot
P-ran\left(  V_{-}\right)  \cdot H-ran\left(  V_{-}\right)  \cdot Q+ran\left(
V_{+}\right)  \cdot Q\\
&  =2\cdot4-2\cdot1-2\cdot2+1\cdot2=4
\end{align*}
and furthemore%

\[
\chi_{S-\exp}=ran\left(  V_{+}\right)  _{S-\exp}-ran\left(  V_{-}\right)
_{S-\exp}=5-1=1~.
\]

On the other hand, using the equation for $\chi_{S-\exp}$ in the presence of
$\bar{\lambda}=0$ elements, it is found%

\[
\chi_{S-\exp}=\chi\cdot\left(  P-H-2Q\right)  =-1\cdot\left(  4-1-4\right)
=1
\]

These results establish the conditions to enable two Lie algebras can be
obtained one from the other, by the $S$-expansion procedure.

If there are no negative or null eigenvalues, we will have%

\begin{align*}
\chi_{S-\exp}  &  =ran\left(  V_{+}\right)  \cdot P-ran\left(  V_{-}\right)
\cdot P\\
&  =\chi\cdot P~,
\end{align*}
on the other hand,%

\begin{align}
\chi_{S-\exp}  &  =\chi\cdot\left(  P-H-2Q\right) \nonumber\\
If~~H  &  =Q=0\text{, \ }\chi_{S-\exp}=\chi\cdot P\nonumber
\end{align}

\subsection{\textbf{Characteristics of the term} $K_{i\gamma}^{\delta
}K_{j\delta}^{\gamma}$}

Since $K_{\alpha\beta}^{\gamma}\in\left\{  0,1\right\}  $ we have $K_{i\gamma
}^{\delta}K_{j\delta}^{\gamma}\in\left\{  0,1,2,\#=P\right\}  $ ($\#$ denoting
cardinality) with indices $i$ and $j$ in the set $\left\{  1,2,\ldots
,P\right\}  $. This statement can be corroborated by the following example:
Consider the product $K_{i\gamma}^{\delta}K_{j\delta}^{\gamma}$ for the case
$i~,~j=1,1$:%

\begin{align}
&  K_{1\gamma}^{\delta}K_{1\delta}^{\gamma}=K_{11}^{\delta}K_{1\delta}%
^{1}+K_{12}^{\delta}K_{1\delta}^{2}+\cdots+K_{1P}^{\delta}K_{1\delta}%
^{P}\nonumber\\
&  =K_{11}^{1}K_{11}^{1}+K_{11}^{2}K_{12}^{1}+\cdots+K_{11}^{P}K_{1P}%
^{1}\nonumber\\
&  +K_{12}^{1}K_{11}^{2}+K_{12}^{2}K_{12}^{2}+\cdots+K_{12}^{P}K_{1P}^{2}\\
&  \vdots\nonumber\\
&  +K_{1P}^{1}K_{11}^{P}+K_{13}^{2}K_{12}^{3}+\cdots+K_{1P}^{P}K_{1P}%
^{P}~.\nonumber
\end{align}

Note that not all summands may be different from zero. If this happens could
not been univocally defined the composition law semigroup. For example if we
assume that%

\begin{equation}
K_{13}^{1}K_{11}^{3}\neq0\text{ \ ; \ }K_{13}^{2}K_{12}^{3}\neq0\text{ \ ;
}\cdots\text{\ }K_{13}^{P}K_{1P}^{3}\neq0
\end{equation}
we would come to a contradiction. Indeed, if $K_{13}^{1}K_{11}^{3}\neq0$ then
$K_{13}^{1}\neq0$ $\ \ \wedge\ \ K_{11}^{3}\neq0$, so that the products
$s_{1}\diamondsuit s_{3}$ and $s_{1}\diamondsuit s_{1}$ are defined as
$s_{1}\diamondsuit s_{3}=s_{1}$ and $s_{1}\diamondsuit s_{1}=s_{3}$. This
means that due to this choice $K_{13}^{2}K_{12}^{3}=0,~\cdots,$ \ $K_{13}%
^{P}K_{1P}^{3}=0$.

Since this is true for any pair of indices $i,j$ we have that the maximum
value of $K_{i\gamma}^{\delta}K_{j\delta}^{\gamma}$ is equal to the number of
elements that form the semigroup. So we have:%

\begin{align*}
K_{1\gamma}^{\delta}K_{1\delta}^{\gamma}  &  =K_{11}^{\delta}K_{1\delta}%
^{1}+K_{12}^{\delta}K_{1\delta}^{2}+K_{13}^{\delta}K_{1\delta}^{3}\\
&  =K_{11}^{1}K_{11}^{1}+K_{11}^{2}K_{12}^{1}+\cdots+K_{11}^{P}K_{1P}^{1}\\
&  +K_{12}^{1}K_{11}^{2}+K_{12}^{2}K_{12}^{2}+\cdots+K_{12}^{P}K_{1P}^{2}\\
&  \vdots\\
&  +K_{13}^{1}K_{11}^{3}+K_{13}^{2}K_{12}^{3}+\cdots+K_{1P}^{P}K_{1P}^{P}\\
&  =1+1+\cdots+1=P\text{ \ }\forall\text{ }\left\{  i,j\right\}  \in\left\{
1,2,\ldots,P\right\}
\end{align*}

This has the consequence that all $KK$ pairs that appear in the expanded
metric multiplied by the coefficients associated with the original algebra,
have at most a magnitud equal to "order" $P$ of the semigroup.

It should be noted that if the semigroup has$~P$ elements, then there must
exist a "$z$" number of $K$-selectors that satisfy the condition $K_{i\gamma
}^{\delta}\neq0$ for some combination of their indices, according to the
following table:%

\begin{align}
P_{1}  &  =1\nonumber\\
z  &  =1\nonumber\\
P_{2}  &  =2\nonumber\\
z  &  =P_{1}+P_{2}=1+2\nonumber\\
P_{3}  &  =3\nonumber\\
z  &  =P_{1}+P_{2}+P_{3}=1+2+3\nonumber\\
&  \vdots\nonumber\\
P_{P}  &  =P\nonumber\\
z  &  =p_{1}+p_{2}+p_{3}+\cdots+p_{p}=1+2+3+\cdots+P\nonumber\\
&  =\sum_{i=1}^{P}P_{i}=\sum_{i=1}^{P}i=\frac{i\left(  i+1\right)  }{2}%
\end{align}

it%
\'{}%
s interesting that this is just the Faulhaber%
\'{}%
s formula $\sum_{i=1}^{P}i^{n}$ for $n=1$. That expresses the well known
"triangular numbers" showed in ref. \cite{Faulhaber}

\subsection{\textbf{Vector Magnitude}}

An interesting consequence of the above property is that $K_{\alpha\gamma
}^{\delta}K_{\alpha\delta}^{\gamma}\in\left\{  0,1,2,\cdots,P\right\}  $
affects the measurement of the length of the original basis vectors. Indeed,%

\begin{align}
\left\Vert X_{\Phi}\right\Vert  &  =\sqrt{\left(  X_{\Phi},X_{\Phi}\right)
_{S-\exp}}=\sqrt{tr\left(  R\left(  X_{\Phi}\right)  R\left(  X_{\Phi}\right)
\right)  }\nonumber\\
&  =\sqrt{R\left(  X_{\Phi}\right)  _{\Omega}^{\Theta}R\left(  X_{\Phi
}\right)  _{\Theta}^{\Omega}}=\sqrt{\left(  C_{\Phi}\right)  _{\Omega}%
^{\Theta}\left(  C_{\Phi}\right)  _{\Theta}^{\Omega}}=\sqrt{\left(  C_{\left(
\alpha,A\right)  }\right)  _{\left(  \gamma,C\right)  }^{\left(
\delta,D\right)  }\left(  C_{\left(  \alpha,A\right)  }\right)  _{\left(
\delta,D\right)  }^{\left(  \gamma,C\right)  }}\nonumber\\
&  =\sqrt{K_{\alpha\gamma}^{\delta}K_{\alpha\delta}^{\gamma}C_{AC}^{D}%
C_{AD}^{C}}=\sqrt{K_{\alpha\gamma}^{\delta}K_{\alpha\delta}^{\gamma}}%
\sqrt{C_{AC}^{D}C_{AD}^{C}}=\sqrt{K_{\alpha\gamma}^{\delta}K_{\alpha\delta
}^{\gamma}}\sqrt{g_{AA}}=\sqrt{K_{\alpha\gamma}^{\delta}K_{\alpha\delta
}^{\gamma}}\left\Vert X_{A}\right\Vert
\end{align}

\begin{align}
\left\Vert X_{\Phi=\left(  \alpha,A\right)  }\right\Vert  &  =\sqrt
{K_{\alpha\gamma}^{\delta}K_{\alpha\delta}^{\gamma}}\left\Vert X_{A}%
\right\Vert \nonumber\\
&  \in%
\begin{Bmatrix}
\sqrt{0\cdot g_{AA}}\equiv\sqrt{0}\left\Vert X_{A}\right\Vert =g_{\left(
\alpha,A\right)  \left(  \alpha,A\right)  }\\
\sqrt{1\cdot g_{AA}}\equiv\sqrt{1}\left\Vert X_{A}\right\Vert =g_{\left(
\alpha,A\right)  \left(  \alpha,A\right)  }\\
\sqrt{2\cdot g_{AA}}\equiv\sqrt{2}\left\Vert X_{A}\right\Vert =g_{\left(
\alpha,A\right)  \left(  \alpha,A\right)  }\\
\sqrt{3\cdot g_{AA}}\equiv\sqrt{3}\left\Vert X_{A}\right\Vert =g_{\left(
\alpha,A\right)  \left(  \alpha,A\right)  }\\
\vdots\\
\sqrt{P\cdot g_{AA}}\equiv\sqrt{p}\left\Vert X_{A}\right\Vert =g_{\left(
\alpha,A\right)  \left(  \alpha,A\right)  }%
\end{Bmatrix}
\end{align}

or in the general case%

\begin{equation}
\left\Vert V\right\Vert =\sqrt{\left(  v^{\left(  \alpha,A\right)  }\right)
^{2}g_{\left(  \alpha,A\right)  \left(  \alpha,A\right)  }}=v^{\left(
\alpha,A\right)  }\sqrt{K_{\alpha\gamma}^{\delta}K_{\alpha\delta}^{\gamma}%
}\left\Vert X_{A}\right\Vert
\end{equation}

This means that the metric tensor components experience a rescaling. We can
also see that the rescaling of vectors depends on elements of $K_{i\gamma
}^{\delta}K_{j\delta}^{\gamma}\longrightarrow K_{\alpha\gamma}^{\delta
}K_{\alpha\delta}^{\gamma}$, present in the diagonal of the expanded metric tensor.

\subsection{\textbf{Angular spacing between vectors}}

The S-expansion procedure affects the angle between two vectors in the
$V_{+}+V_{-}=\left(  S\otimes\mathcal{G}\right)  /V_{0}\subseteq
S\otimes\mathcal{G}$ space. Indeed,

\bigskip%

\begin{align}
\cos\left(  \theta\right)  _{S-\exp}  &  =\frac{\left(  X_{\Phi},X_{\Theta
}\right)  _{S-\exp}}{\sqrt{\left(  X_{\Phi},X_{\Phi}\right)  _{S-\exp}}%
\sqrt{\left(  X_{\Theta},X_{\Theta}\right)  _{S-\exp}}}\equiv\frac{\left(
X_{\Phi},X_{\Theta}\right)  }{\sqrt{\left(  X_{\Phi},X_{\Phi}\right)  }%
\sqrt{\left(  X_{\Theta},X_{\Theta}\right)  }}\nonumber\\
&  =\frac{tr\left(  R\left(  X_{\Phi}\right)  R\left(  X_{\Theta}\right)
\right)  }{\sqrt{tr\left(  R\left(  X_{\Phi}\right)  R\left(  X_{\Phi}\right)
\right)  }\sqrt{tr\left(  R\left(  X_{\Theta}\right)  R\left(  X_{\Theta
}\right)  \right)  }}\nonumber\\
&  =\frac{R\left(  X_{\Phi}\right)  _{\Omega}^{\Gamma}R\left(  X_{\Theta
}\right)  _{\Gamma}^{\Omega}}{\sqrt{trR\left(  X_{\Phi}\right)  _{\Omega
}^{\Gamma}R\left(  X_{\Phi}\right)  _{\Gamma}^{\Omega}}\sqrt{R\left(
X_{\Theta}\right)  _{\Omega}^{\Gamma}R\left(  X_{\Theta}\right)  _{\Omega
}^{\Gamma}}}\\
&  =\frac{\left(  C_{\Phi}\right)  _{\Omega}^{\Gamma}\left(  C_{\Theta
}\right)  _{\Gamma}^{\Omega}}{\sqrt{\left(  C_{\Phi}\right)  _{\Omega}%
^{\Gamma}\left(  C_{\Phi}\right)  _{\Gamma}^{\Omega}}\sqrt{\left(  C_{\Theta
}\right)  _{\Omega}^{\Gamma}\left(  C_{\Theta}\right)  _{\Omega}^{\Gamma}}%
}\nonumber\\
&  =\frac{\left(  C_{\left(  \alpha,A\right)  }\right)  _{\left(
\gamma,C\right)  }^{\left(  \delta,D\right)  }\left(  C_{\left(
\beta,B\right)  }\right)  _{\left(  \delta,D\right)  }^{\left(  \gamma
,C\right)  }}{\sqrt{\left(  C_{\left(  \alpha,A\right)  }\right)  _{\left(
\gamma,C\right)  }^{\left(  \delta,D\right)  }\left(  C_{\left(
\alpha,A\right)  }\right)  _{\left(  \delta,D\right)  }^{\left(
\gamma,C\right)  }}\sqrt{\left(  C_{\left(  \beta,B\right)  }\right)
_{\left(  \gamma,C\right)  }^{\left(  \delta,D\right)  }\left(  C_{\left(
\beta,B\right)  }\right)  _{\left(  \delta,D\right)  }^{\left(  \gamma
,C\right)  }}}\\
&  =\frac{K_{\alpha\gamma}^{\delta}K_{\beta\delta}^{\gamma}\left(
C_{A}\right)  _{C}^{D}\left(  C_{B}\right)  _{D}^{C}}{\sqrt{K_{\alpha\gamma
}^{\delta}K_{\alpha\delta}^{\gamma}\left(  C_{A}\right)  _{C}^{D}\left(
C_{A}\right)  _{D}^{C}}\sqrt{K_{\beta\gamma}^{\delta}K_{\beta\delta}^{\gamma
}\left(  C_{B}\right)  _{C}^{D}\left(  C_{B}\right)  _{D}^{C}}}\nonumber
\end{align}

\begin{align}
&  =\frac{K_{\alpha\gamma}^{\delta}K_{\beta\delta}^{\gamma}}{\sqrt
{K_{\alpha\gamma}^{\delta}K_{\alpha\delta}^{\gamma}}\sqrt{K_{\beta\gamma
}^{\delta}K_{\beta\delta}^{\gamma}}}\frac{\left(  C_{A}\right)  _{C}%
^{D}\left(  C_{B}\right)  _{D}^{C}}{\sqrt{\left(  C_{A}\right)  _{C}%
^{D}\left(  C_{A}\right)  _{D}^{C}}\sqrt{\left(  C_{B}\right)  _{C}^{D}\left(
C_{B}\right)  _{D}^{C}}}\nonumber\\
\left(  X_{\left(  \alpha,A\right)  },X_{\left(  \beta,B\right)  }\right)   &
=\frac{K_{\alpha\gamma}^{\delta}K_{\beta\delta}^{\gamma}}{\sqrt{K_{\alpha
\gamma}^{\delta}K_{\alpha\delta}^{\gamma}}\sqrt{K_{\beta\gamma}^{\delta
}K_{\beta\delta}^{\gamma}}}\frac{g_{AB}}{\sqrt{g_{AA}}\sqrt{g_{BB}}%
}\nonumber\\
&  =\frac{\sum K_{\alpha\gamma}^{\delta}K_{\beta\delta}^{\gamma}}{\sqrt{\sum
K_{\alpha\gamma}^{\delta}K_{\alpha\delta}^{\gamma}}\sqrt{\sum K_{\beta\gamma
}^{\delta}K_{\beta\delta}^{\gamma}}}\cos\left(  \theta\right)
\end{align}

\bigskip In general we find that,%

\begin{align}
\left(  v^{\left(  \alpha,A\right)  }X_{\left(  \alpha,A\right)  },v^{\left(
\beta,B\right)  }X_{\left(  \beta,B\right)  }\right)   &  =v^{\left(
\alpha,A\right)  }v^{\left(  \beta,B\right)  }\frac{K_{\alpha\gamma}^{\delta
}K_{\beta\delta}^{\gamma}}{\sqrt{K_{\alpha\gamma}^{\delta}K_{\alpha\delta
}^{\gamma}}\sqrt{K_{\beta\gamma}^{\delta}K_{\beta\delta}^{\gamma}}}%
\frac{g_{AB}}{\sqrt{g_{AA}}\sqrt{g_{BB}}}\nonumber\\
&  =v^{\left(  \alpha,A\right)  }v^{\left(  \beta,B\right)  }\frac{\sum
K_{\alpha\gamma}^{\delta}K_{\beta\delta}^{\gamma}}{\sqrt{\sum K_{\alpha\gamma
}^{\delta}K_{\alpha\delta}^{\gamma}}\sqrt{\sum K_{\beta\gamma}^{\delta
}K_{\beta\delta}^{\gamma}}}\cos\left(  \theta\right)
\end{align}

So that%

\begin{align*}
\cos\left(  \Theta\right)   &  =\left(  X_{\left(  \alpha,A\right)
},X_{\left(  \beta,B\right)  }\right)  =\frac{\sum K_{\alpha\gamma}^{\delta
}K_{\beta\delta}^{\gamma}}{\sqrt{K_{\alpha\gamma}^{\delta}K_{\alpha\delta
}^{\gamma}}\sqrt{K_{\beta\gamma}^{\delta}K_{\beta\delta}^{\gamma}}}%
\frac{g_{AB}}{\sqrt{g_{AA}}\sqrt{g_{BB}}}\\
&  =\frac{\sum K_{\alpha\gamma}^{\delta}K_{\beta\delta}^{\gamma}}{\sqrt{\sum
K_{\alpha\gamma}^{\delta}K_{\alpha\delta}^{\gamma}}\sqrt{\sum K_{\beta\gamma
}^{\delta}K_{\beta\delta}^{\gamma}}}\cos\left(  \theta\right)  =n\in%
\mathbb{R}%
\end{align*}

These results lead to define a number denoted as $\Delta$ which depends on the
composition law of the semigroup. In fact the number $\Delta$ is defined as%

\[
\Delta\equiv\left\{  \frac{\left[  0,1,2,3,\cdots,P\right]  _{i,j}}%
{\sqrt{\left[  0,1,2,3,\cdots,P\right]  _{i,i}}\sqrt{\left[  0,1,2,3,\cdots
,P\right]  _{j,j}}}\right\}  \in%
\mathbb{R}
^{+},
\]
from where we can see that $\Delta$ depends on the composition law semigroup
and that is a real number. These results also lead to an important condition
for the diagonalization of the $S$-expanded metric depending on the internal
law of the semigroup. In fact the metric will be diagonal if:%

\begin{equation}
\sum K_{i\gamma}^{\delta}K_{j\delta}^{\gamma}=0
\end{equation}

for $i\not =j$.

Although the amount $KK$ is positive semidefinite, it is possible that there
is a base change such that the diagonal elements $K_{i\gamma}^{\delta
}K_{i\delta}^{\gamma}$ change of sign.

However this does not affect the good performance of the function $\cos\left(
\Theta\right)  $ in a diagonal base.

Since the function $\cos\left(  \Theta\right)  $ is bounded in a continuous
interval $\left[  -1,1\right]  $, we have some values for $\Delta$ that are
prohibited. In fact for an arbitrary value of $\theta$ angle, $i~,~j$ fixed we
could have%

\begin{equation}
\cos\left(  \Theta\right)  =\left(  X_{\left(  i,A\right)  },X_{\left(
j,B\right)  }\right)  =\frac{\sum K_{i\gamma}^{\delta}K_{j\delta}^{\gamma}%
}{\sqrt{\sum K_{i\gamma}^{\delta}K_{i\delta}^{\gamma}}\sqrt{\sum K_{j\gamma
}^{\delta}K_{j\delta}^{\gamma}}}\cos\left(  \theta\right)  \neq\frac{0}%
{0},\nonumber
\end{equation}
from where we can see that the following conditions%

\begin{align}
\forall\ K_{i\gamma}^{\delta}K_{j\delta}^{\gamma}  &  =0,\ K_{i\gamma}%
^{\delta}K_{i\delta}^{\gamma}\cdot K_{j\gamma}^{\delta}K_{j\delta}^{\gamma
}\neq0\\
\forall\ K_{i\gamma}^{\delta}K_{j\delta}^{\gamma}  &  \neq0,\ K_{i\gamma
}^{\delta}K_{i\delta}^{\gamma}\cdot K_{j\gamma}^{\delta}K_{j\delta}^{\gamma
}\neq0
\end{align}
must be fulfilled.

That is, in every case of the configurations for the $S-$expanded metric must
be met that $K_{i\gamma}^{\delta}K_{i\delta}^{\gamma}\neq0$ in order to
prevent that the angular separation between vectors of this diagonal basis is
ill-defined when we make the product $S\otimes\mathcal{G}$.

Since the function $\cos\left(  \Theta\right)  $ between two different basis
vectors of these must be zero and not $0/0$ or $\infty$, it is necessary that
the diagonal elements $K_{i\gamma}^{\delta}K_{i\delta}^{\gamma}$ are different
from zero, so that the angular separation of the new basis vectors is well
defined. For example for the case $P=3$, it is found%

\begin{align}
\cos\left(  \Theta\right)   &  =\left(  X_{\left(  1,A\right)  },X_{\left(
3,B\right)  }\right)  =\frac{\sum K_{1\gamma}^{\delta}K_{3\delta}^{\gamma}%
}{\sqrt{\sum K_{1\gamma}^{\delta}K_{1\delta}^{\gamma}}\sqrt{\sum K_{3\gamma
}^{\delta}K_{3\delta}^{\gamma}}}\cos\left(  \theta\right)  _{A,B}\nonumber\\
\forall\text{ \ \ }K_{1\gamma}^{\delta}K_{1\delta}^{\gamma}\cdot K_{3\gamma
}^{\delta}K_{3\delta}^{\gamma}  &  \neq0~.
\end{align}

If $X_{A}$ and $X_{B}$ are ortogonal in the original algebra and then $\sum
K_{1\gamma}^{\delta}K_{3\delta}^{\gamma}\cdot\cos\left(  \theta\right)
_{A,B}$\ $=0$ and we could have, for example%

\[
\frac{\sum K_{1\gamma}^{\delta}K_{3\delta}^{\gamma}}{\sqrt{\sum K_{1\gamma
}^{\delta}K_{1\delta}^{\gamma}}\sqrt{\sum K_{3\gamma}^{\delta}K_{3\delta
}^{\gamma}}}\cos\left(  \theta\right)  _{A,B}=0
\]

But if further the term $\sum K_{1\gamma}^{\delta}K_{1\delta}^{\gamma}$ or the
term $\sum K_{3\gamma}^{\delta}K_{3\delta}^{\gamma}$ is zero we have a
mathematical problem regardless the value of $\sum K_{1\gamma}^{\delta
}K_{3\delta}^{\gamma}$, then it%
\'{}%
s always necessary a $M_{K}$ matrix with the form%

\begin{equation}
M_{K}\equiv%
\begin{pmatrix}
K_{1\gamma}^{\delta}K_{1\delta}^{\gamma}\neq0 & \ast & \ast\\
\ast & \ast & \ast\\
\ast & \ast & K_{3\gamma}^{\delta}K_{3\delta}^{\gamma}\neq0
\end{pmatrix}
.
\end{equation}
The boxes containing the symbol $\ast$ may be zero simultaneously or
separately, for some choice of semigroup.

For other cases it is found%

\begin{align}
M_{K}  &  \equiv%
\begin{pmatrix}
K_{1\gamma}^{\delta}K_{1\delta}^{\gamma}\neq0 & \ast & \ast\\
\ast & K_{2\gamma}^{\delta}K_{2\delta}^{\gamma}\neq0 & \ast\\
\ast & \ast & \ast
\end{pmatrix}
\\
M_{K}  &  \equiv%
\begin{pmatrix}
\ast & \ast & \ast\\
\ast & K_{2\gamma}^{\delta}K_{2\delta}^{\gamma}\neq0 & \ast\\
\ast & \ast & K_{3\gamma}^{\delta}K_{3\delta}^{\gamma}\neq0
\end{pmatrix}
\end{align}

\bigskip

In summary we can say that both the magnitude of the vectors and the angle
between them are strongly affected\ by $S$-expansion process. In more precise
way they are affected by the composition law of the semigroup codified in the
$K$-selectors.

It should be noted that in the last calculation, both indices of the original
algebra as semigroup indices $i,j$ are fixed. This means that the change in
the magnitude or the angle $\theta$ depends on the sum of $\gamma,\delta$
indices. So if we want to change these geometric properties is necessary to
impose conditions on the semigroup from this sum.

\section{\textbf{\ The semigroup: example }$\mathbf{so(4)}$ \textbf{from}
$\mathbf{so(3)}$}

So far we have studied the effects it produces, the process of expansion, on
the geometry of the manifold of the original Lie group. In particular, we have
considered the effects on the metric of the manifold of Lie group that lead us
to the metric of a new Lie group. The result of this Study allows to determine
the geometrical role of the semigroup and its composition law. In this section
is outlined, via an example, a method for determining the semigroup, which
would provide a Lie algebra from another. This problem was recently addressed,
from a slightly different viewpoint, in Refs. \cite{merino}, \cite{merino1}.

\subsection{\textbf{Geometrical considerations}}

Consider first the case in wich $M_{K}$ matrices have no negative or zero
eigenvalues. This means that our attention will focus on the coefficient $"P"
$ with the law of composition $"\diamondsuit"$ semigroup.

In this case obtaining an algebra from another shall be subject to following conditions:%

\begin{align}
&  ran\left(  V_{\pm}\right)  \underset{S-\exp}{\longrightarrow}ran\left(
V_{\pm}\right)  \cdot P\tag{$A.1$}\\
&  ran\left(  \mathcal{G}\right)  \underset{S-\exp}{\longrightarrow}ran\left(
\mathcal{G}\right)  \cdot P\tag{$B.1$}\\
&  \chi\underset{S-\exp}{\longrightarrow}\chi\cdot P\text{ .} \tag{$C.1$}%
\end{align}

For clarity, consider the example of obtaining $so\left(  4\right)  $ from
$so\left(  3\right)  $. In this case one has \ %

\begin{align*}
ran\left(  so\left(  3\right)  \right)   &  =3\\
\chi_{so\left(  3\right)  }  &  =-3\\
ran\left(  so\left(  4\right)  \right)   &  =6\\
\chi_{so\left(  4\right)  }  &  =-6
\end{align*}

\begin{align*}
ran\left(  V_{-}\right)  _{so\left(  3\right)  }  &  =3\\
ran\left(  V_{+}\right)  _{so\left(  3\right)  }  &  =0\\
ran\left(  V_{-}\right)  _{so\left(  4\right)  }  &  =6\\
ran\left(  V_{+}\right)  _{so\left(  4\right)  }  &  =0~.
\end{align*}

This information allows to determine the number of elements that must have the
semigroup connecting such algebras.

~~~~ The character of the expanded algebra is given, as we have seen, by%

\begin{align*}
\chi_{S-\exp}  &  =ran\left(  V_{+}\right)  \cdot\left(  P-H-Q\right)
+ran\left(  V_{-}\right)  \cdot Q-\left[  ran\left(  V_{-}\right)
\cdot\left(  P-H-Q\right)  +ran\left(  V_{+}\right)  \cdot Q\right] \\
&  =-ran\left(  V_{-}\right)  \cdot\left(  P-H-Q\right)  =-3\left(
P-H-Q\right)
\end{align*}
where we see that%

\begin{align*}
-6  &  =-3\left(  P-0-0\right) \\
2  &  =P-0-0\\
2  &  =P
\end{align*}

The ranks of $\left(  V_{-}\right)  _{S-\exp}$ and $\mathcal{G}_{S-\exp}$ are
given by%

\begin{align*}
ran\left(  V_{-}\right)  _{S-\exp}  &  =ran\left(  V_{-}\right)  \cdot\left(
P-H-Q\right)  +ran\left(  V_{+}\right)  \cdot Q\\
6  &  =3\cdot\left(  P-H-Q\right) \\
2  &  =P-0-0\\
ran\left(  \mathcal{G}\right)  _{S-\exp}  &  =ran\left(  \mathcal{G}\right)
\cdot\left(  P-0-0\right) \\
2  &  =P
\end{align*}

This is because the character $so\left(  n\right)  $ is given by
$\chi=-ran\left(  g_{ab}\right)  _{so\left(  n\right)  }$, because all the
generators are compact. So we have that the number of elements is
characterized by $P=2$ , $H=Q=0$ and denoted by $S_{P,H,Q}\longrightarrow
S_{2,0,0}$, i.e., the semigroup may be a semigroup of two elements whose
matrix $M_{K}$ does not have eigenvalues zero and negative eigenvalues.

Now consider the case where there are negative eigenvalues and $ran\left(
V_{\pm}\right)  \neq0$. In this case it is required%

\begin{align}
&  ran\left(  V_{\pm}\right)  \underset{S-\exp}{\longrightarrow}ran\left(
V_{\pm}\right)  \cdot\left(  P-Q\right)  +ran\left(  V_{\mp}\right)  \cdot
Q\tag{$A.2$}\\
&  ran\left(  \mathcal{G}\right)  \underset{S-\exp}{\longrightarrow}ran\left(
\mathcal{G}\right)  \cdot P\tag{$B.2$}\\
&  \chi\underset{S-\exp}{\longrightarrow}\chi\cdot\left(  P-2Q\right)
\tag{$C.2$}%
\end{align}

and for the more general case, $rang\left(  M_{K}\right)  <P$, and again
($ran\left(  V_{\pm}\right)  \neq0$)%

\begin{align}
&  ran\left(  V_{\pm}\right)  \underset{S-\exp}{\longrightarrow}ran\left(
V_{\pm}\right)  \cdot\left(  P-H-Q\right)  +ran\left(  V_{\mp}\right)  \cdot
Q\tag{$A.3$}\\
&  ran\left(  \mathcal{G}\right)  \underset{S-\exp}{\longrightarrow}ran\left(
\mathcal{G}\right)  \cdot\left(  P-H\right) \tag{$B.3$}\\
&  \chi\underset{S-\exp}{\longrightarrow}\chi\cdot\left(  P-H-2Q\right)
~\text{.} \tag{$C.3$}%
\end{align}

This ensures that two Lie algebras $A$ and $B$ could be related by
$S$-expansion, if the total of the above conditions are satisfy.

It should be noted that the above equations are not all indepedientes, because
the equation for character $\chi$ is constructed based on the other. So we
have that the equations%

\begin{align}
&  ran\left(  V_{\pm}\right)  \underset{S-\exp}{\longrightarrow}ran\left(
V_{\pm}\right)  \cdot\left(  P-H-Q\right)  +ran\left(  V_{\mp}\right)  \cdot
Q\\
&  ran\left(  \mathcal{G}\right)  \underset{S-\exp}{\longrightarrow}ran\left(
\mathcal{G}\right)  \cdot\left(  P-H\right)
\end{align}
are independent. This means it is possible that two Lie algebras could be
linked for more than a semigroup in the event that $P\,,H,Q$ are nonzero and
belong to $%
\mathbb{N}
^{\ast}$.

Consider now the conditions that lead to determining the semigroup, i.e. to
the determination of the elements and its composition law.

\subsection{\textbf{Conditions on the semigroup}}

We have established the conditions on the intrinsic geometry generating
conditions on the semigroup. Equivalently, the conditions on the metric leads
to a set of values for the different elements $K_{i\gamma}^{\delta}K_{j\delta
}^{\gamma}$. To clarify the idea consider a semigroup of two elements having a
matrix $M_{K}$ given by%

\begin{equation}
M_{K}=%
\begin{pmatrix}
K_{1\gamma}^{\delta}K_{1\delta}^{\gamma} & K_{1\gamma}^{\delta}K_{2\delta
}^{\gamma}\\
K_{2\gamma}^{\delta}K_{1\delta}^{\gamma} & K_{2\gamma}^{\delta}K_{2\delta
}^{\gamma}%
\end{pmatrix}
\equiv%
\begin{pmatrix}
a & b\\
b & c
\end{pmatrix}
\end{equation}
whose eigenvalues are given by%

\begin{align}
\lambda_{1}  &  =\frac{1}{2}a+\frac{1}{2}c+\frac{1}{2}\sqrt{a^{2}%
-2ac+4b^{2}+c^{2}}\nonumber\\
&  =\frac{1}{2}\left(  K_{1\gamma}^{\delta}K_{1\delta}^{\gamma}+K_{2\gamma
}^{\delta}K_{2\delta}^{\gamma}+\sqrt{\left(  K_{1\gamma}^{\delta}K_{1\delta
}^{\gamma}\right)  ^{2}-2\left(  K_{1\gamma}^{\delta}K_{1\delta}^{\gamma
}\right)  \left(  K_{2\gamma}^{\delta}K_{2\delta}^{\gamma}\right)  +4\left(
K_{1\gamma}^{\delta}K_{2\delta}^{\gamma}\right)  ^{2}+\left(  K_{2\gamma
}^{\delta}K_{2\delta}^{\gamma}\right)  ^{2}}\right) \nonumber\\
\lambda_{2}  &  =\frac{1}{2}a+\frac{1}{2}c-\allowbreak\frac{1}{2}\sqrt
{a^{2}-2ac+4b^{2}+c^{2}}\\
&  =\frac{1}{2}\left(  K_{1\gamma}^{\delta}K_{1\delta}^{\gamma}+K_{2\gamma
}^{\delta}K_{2\delta}^{\gamma}+\sqrt{\left(  K_{1\gamma}^{\delta}K_{1\delta
}^{\gamma}\right)  ^{2}-2\left(  K_{1\gamma}^{\delta}K_{1\delta}^{\gamma
}\right)  \left(  K_{2\gamma}^{\delta}K_{2\delta}^{\gamma}\right)  +4\left(
K_{1\gamma}^{\delta}K_{2\delta}^{\gamma}\right)  ^{2}+\left(  K_{2\gamma
}^{\delta}K_{2\delta}^{\gamma}\right)  ^{2}}\right) \nonumber
\end{align}

Since $M_{K}$ is a symmetric matrix, its eiegenvalues $\lambda_{1,2}\in%
\mathbb{R}
$, and therefore%

\begin{align*}
\left(  K_{1\gamma}^{\delta}K_{1\delta}^{\gamma}\right)  ^{2}-2\left(
K_{1\gamma}^{\delta}K_{1\delta}^{\gamma}\right)  \left(  K_{2\gamma}^{\delta
}K_{2\delta}^{\gamma}\right)  +4\left(  K_{1\gamma}^{\delta}K_{2\delta
}^{\gamma}\right)  ^{2}+\left(  K_{2\gamma}^{\delta}K_{2\delta}^{\gamma
}\right)  ^{2}  &  \geq0\\
\left(  K_{1\gamma}^{\delta}K_{1\delta}^{\gamma}-K_{2\gamma}^{\delta
}K_{2\delta}^{\gamma}\right)  ^{2}+4\left(  K_{1\gamma}^{\delta}K_{2\delta
}^{\gamma}\right)  ^{2}  &  \geq0~.
\end{align*}

Since $\lambda_{2}$ can accept negative or zero values we have%

\begin{align*}
\frac{1}{2}a+\frac{1}{2}c-\allowbreak\frac{1}{2}\sqrt{a^{2}-2ac+4b^{2}+c^{2}}
&  \leq0\\
ac  &  \leq b^{2}\\
\left(  K_{1\gamma}^{\delta}K_{1\delta}^{\gamma}\right)  \left(  K_{2\gamma
}^{\delta}K_{2\delta}^{\gamma}\right)   &  \leq\left(  K_{1\gamma}^{\delta
}K_{2\delta}^{\gamma}\right)  ^{2}~.
\end{align*}

This result allows to modify the $\lambda_{1,2}$ values according to
relationships between $K_{i\gamma}^{\delta}K_{j\delta}^{\gamma}$. Since: $(i)$
each $K_{i\gamma}^{\delta}K_{j\delta}^{\gamma}$ object takes values ranging
between $0$ and $P$ (or between $1$ and $P$) , which in this case are bounded
by $\left\{  0,1,2\right\}  $ and $(ii)$ the angular separation (orthogonality
metric if diagonal) must be well defined in space $S\times\mathcal{G}$
irrespective of the chosen base; it is necessary that all elements of type
$K_{i\gamma}^{\delta}K_{i\delta}^{\gamma}$ are nonzero. So we have to
$\lambda_{2}<0$%

\begin{align*}
\left(  K_{1\gamma}^{\delta}K_{1\delta}^{\gamma}\right)  \left(  K_{2\gamma
}^{\delta}K_{2\delta}^{\gamma}\right)   &  <\left(  K_{1\gamma}^{\delta
}K_{2\delta}^{\gamma}\right)  ^{2}\\
1\cdot1  &  <2^{2}=4\\
1\cdot2  &  <2^{2}=4\\
2\cdot1  &  <2^{2}=4
\end{align*}
for the case of a null eigenvector:%

\[
\left(  K_{1\gamma}^{\delta}K_{1\delta}^{\gamma}\right)  \left(  K_{2\gamma
}^{\delta}K_{2\delta}^{\gamma}\right)  =\left(  K_{1\gamma}^{\delta}%
K_{2\delta}^{\gamma}\right)  ^{2}%
\]

\[
1\cdot1=1^{2}=1
\]

or%

\[
2\cdot2=2^{2}=4
\]

Note that this condition is equivalent to $\det\left(  M_{K}\right)  =0$. For
a semigroup of order three $M_{K}$ have the form%

\[%
\begin{pmatrix}
K_{1\gamma}^{\delta}K_{1\delta}^{\gamma} & K_{1\gamma}^{\delta}K_{2\delta
}^{\gamma} & K_{1\gamma}^{\delta}K_{3\delta}^{\gamma}\\
K_{2\gamma}^{\delta}K_{1\delta}^{\gamma} & K_{2\gamma}^{\delta}K_{2\delta
}^{\gamma} & K_{2\gamma}^{\delta}K_{3\delta}^{\gamma}\\
K_{3\gamma}^{\delta}K_{1\delta}^{\gamma} & K_{3\gamma}^{\delta}K_{2\delta
}^{\gamma} & K_{3\gamma}^{\delta}K_{3\delta}^{\gamma}%
\end{pmatrix}
\equiv%
\begin{pmatrix}
a & b & c\\
b & d & e\\
c & e & f
\end{pmatrix}
\]
whose characteristic polynomial is%

\begin{equation}
P\left(  X\right)  =X^{3}+\left(  -a-d-f\right)  X^{2}+\left(  -b^{2}%
-c^{2}+f\left(  a+d\right)  +ad-e^{2}\right)  \allowbreak X+\left(
2ebc-fb^{2}-dc^{2}-ae^{2}+adf\right)  =0~.
\end{equation}

So that

$a)$ For $H=1$, we find%

\begin{equation}
-fb^{2}+2ebc-dc^{2}-ae^{2}+adf=0
\end{equation}

therefore there is at least a null root for the polynomial $P\left(  X\right)
$.

$b)~$For $H=2,$ we have%

\begin{align}
-fb^{2}+2ebc-dc^{2}-ae^{2}+adf  &  =0\tag{$I$}\\
-b^{2}-c^{2}+f\left(  a+d\right)  +ad-e^{2}  &  =0 \tag{$II$}%
\end{align}
which produces at least two null roots in polynomial $P\left(  X\right)  $.

$3)~$For $H=3$%

\begin{align}
-fb^{2}+2ebc-dc^{2}-ae^{2}+adf  &  =0\tag{$I$}\\
-b^{2}-c^{2}+f\left(  a+d\right)  +ad-e^{2}  &  =0\tag{$II$}\\
a+d+f  &  =0 \tag{$III$}%
\end{align}
which produces at least three null roots in polynomial $P\left(  X\right)  $.

This allows to obtain conditions on $K_{i\gamma}^{\delta}K_{j\delta}^{\gamma}$
objects so that the semigroup and its composition law appear as a natural
consequence of the loss of semisimplicity of the expanded algebra to $H\neq0$

Let us see how determine the composition law semigroup $S_{P,H,Q}$.

\subsection{\textbf{Case of a semigroup of order 2}}

Consider an abelian finite and arbitrary semigroup of two elements%

\begin{equation}%
\begin{tabular}
[c]{ccc}%
$\diamondsuit$ & $\lambda_{1}$ & $\lambda_{2}$\\
$\lambda_{1}$ & $\ast$ & $\ast$\\
$\lambda_{2}$ & $\ast$ & $\ast$%
\end{tabular}
~.
\end{equation}

If $Q=1$ we find%

\[
\left(  K_{1\gamma}^{\delta}K_{1\delta}^{\gamma}\right)  \left(  K_{2\gamma
}^{\delta}K_{2\delta}^{\gamma}\right)  <\left(  K_{1\gamma}^{\delta}%
K_{2\delta}^{\gamma}\right)  ^{2}~.
\]

In fact%

\[
K_{i\gamma}^{\delta}K_{j\delta}^{\gamma}=\left\{  K_{1\gamma}^{\delta
}K_{1\delta}^{\gamma},K_{1\gamma}^{\delta}K_{2\delta}^{\gamma},K_{2\gamma
}^{\delta}K_{2\delta}^{\gamma}\right\}
\]

\begin{align*}
K_{1\gamma}^{\delta}K_{1\delta}^{\gamma}  &  =K_{11}^{\delta}K_{1\delta}%
^{1}+K_{12}^{\delta}K_{1\delta}^{2}\\
&  =K_{11}^{1}K_{11}^{1}+K_{11}^{2}K_{12}^{1}+K_{12}^{1}K_{11}^{2}+K_{12}%
^{2}K_{12}^{2}%
\end{align*}

\begin{align*}
K_{2\gamma}^{\delta}K_{2\delta}^{\gamma}  &  =K_{21}^{\delta}K_{2\delta}%
^{1}+K_{22}^{\delta}K_{2\delta}^{2}\\
&  =K_{21}^{1}K_{21}^{1}+K_{21}^{2}K_{22}^{1}+K_{22}^{1}K_{21}^{2}+K_{22}%
^{2}K_{22}^{2}%
\end{align*}

\begin{align*}
K_{1\gamma}^{\delta}K_{2\delta}^{\gamma}  &  =K_{11}^{\delta}K_{2\delta}%
^{1}+K_{12}^{\delta}K_{2\delta}^{2}\\
&  =K_{11}^{1}K_{21}^{1}+K_{11}^{2}K_{22}^{1}+K_{12}^{1}K_{21}^{2}+K_{12}%
^{2}K_{22}^{2}~.
\end{align*}

This allows to establish a restriction on the value of $Q$ using%

\begin{align*}
K_{1\gamma}^{\delta}K_{1\delta}^{\gamma}  &  \in\left\{  1,2\right\} \\
K_{1\gamma}^{\delta}K_{2\delta}^{\gamma}  &  \in\left\{  0,1,2\right\} \\
K_{2\gamma}^{\delta}K_{2\delta}^{\gamma}  &  \in\left\{  1,2\right\}
\end{align*}

and given that%

\begin{align}
K_{12}^{1}K_{21}^{2}  &  =0\tag{$a$}\\
K_{11}^{1}K_{21}^{1}  &  \neq0\text{ \ \ }\vee\text{ \ \ }K_{12}^{2}K_{22}%
^{2}\neq0 \tag{$b$}%
\end{align}
we have%

\begin{equation}
b=K_{1\gamma}^{\delta}K_{2\delta}^{\gamma}\in\left\{  0,1\right\}  ~.
\end{equation}

This means that the condition is not satisfied for $a\cdot c<b^{2},$ when
$a\neq0\neq c$. However this condition is satisfied when one of the diagonal
elements are zero. So, we have%

\[
a\cdot c<b^{2}%
\]
leads to%

\begin{align*}
1\cdot0  &  <1\\
0\cdot1  &  <1\\
2\cdot0  &  <1\\
0\cdot2  &  <1~.
\end{align*}

If we consider the $K_{1\gamma}^{\delta}K_{1\delta}^{\gamma}$ object,%

\begin{align}
K_{1\gamma}^{\delta}K_{1\delta}^{\gamma}  &  =K_{11}^{\delta}K_{1\delta}%
^{1}+K_{12}^{\delta}K_{1\delta}^{2}\tag{$a$}\\
&  =\left(  K_{11}^{1}K_{11}^{1}=0\right)  +\left(  K_{11}^{2}K_{12}^{1}%
\neq0\right)  +\left(  K_{12}^{1}K_{11}^{2}\neq0\right)  +\left(  K_{12}%
^{2}K_{12}^{2}=0\right) \nonumber
\end{align}

or%

\begin{align}
K_{1\gamma}^{\delta}K_{1\delta}^{\gamma}  &  =K_{11}^{\delta}K_{1\delta}%
^{1}+K_{12}^{\delta}K_{1\delta}^{2}\tag{$b$}\\
&  =\left(  K_{11}^{1}K_{11}^{1}\neq0\right)  +\left(  K_{11}^{2}K_{12}%
^{1}=0\right)  +\left(  K_{12}^{1}K_{11}^{2}=0\right)  +\left(  K_{12}%
^{2}K_{12}^{2}\neq0\right) \nonumber
\end{align}

This is not true because the law of composition semigroup is univocally
defined for each pair of elements. That is, it is not possible to find a
semigroup of order two that allows us to exchange parts of space $V_{+}$ with
parts of space $V_{-}$. So we can say that "it is impossible to obtain
$Q\neq0$ using a semigroup $S$ of order $2$".

Consider now $H\in\left\{  0,1,2\right\}  $. Using the equality%

\begin{align*}
\left(  K_{1\gamma}^{\delta}K_{1\delta}^{\gamma}\right)  \left(  K_{2\gamma
}^{\delta}K_{2\delta}^{\gamma}\right)   &  =\left(  K_{1\gamma}^{\delta
}K_{2\delta}^{\gamma}\right)  ^{2}\\
a\cdot c  &  =b^{2}\\
&  for\\
1\cdot1  &  =1\\
2\cdot2  &  =2^{2}%
\end{align*}
This means that $K_{1\gamma}^{\delta}K_{2\delta}^{\gamma}<2$, because there is
no way that more than two terms of this element is not null. So that%

\begin{align}
b  &  <2,\text{ \ i.e., }b\in\left\{  0,1\right\} \nonumber\\
ac  &  =b^{2}\nonumber\\
1\cdot1  &  =1
\end{align}

and%

\begin{align*}
1\cdot0  &  =0\cdot1=0\\
2\cdot0  &  =0\cdot2=0
\end{align*}
But the last are forbidden. So we have that, for $H=1$, the $K_{i\gamma
}^{\delta}K_{j\delta}^{\gamma}$ matrix takes the form%

\begin{align*}
\left(  K_{i\gamma}^{\delta}K_{j\delta}^{\gamma}\right)   &  =%
\begin{pmatrix}
1 & 1\\
1 & 1
\end{pmatrix}
\\
_{d}\left(  K_{i\gamma}^{\delta}K_{j\delta}^{\gamma}\right)   &  =%
\begin{pmatrix}
2 & 0\\
0 & 0
\end{pmatrix}
.
\end{align*}

To calculate the semigroup corresponding to this matrix $M_{K}$, we use%

\begin{align}
K_{1\gamma}^{\delta}K_{1\delta}^{\gamma}  &  =1\nonumber\\
K_{1\gamma}^{\delta}K_{2\delta}^{\gamma}  &  =1=K_{2\gamma}^{\delta}%
K_{1\delta}^{\gamma}\nonumber\\
K_{2\gamma}^{\delta}K_{2\delta}^{\gamma}  &  =1
\end{align}

We begin with the first double sum on a possible semigroup of the type
$S_{2,1,0}$%

\begin{align*}
K_{1\gamma}^{\delta}K_{1\delta}^{\gamma}  &  =K_{11}^{\delta}K_{1\delta}%
^{1}+K_{12}^{\delta}K_{1\delta}^{2}\\
&  =K_{11}^{1}K_{11}^{1}+K_{11}^{2}K_{12}^{1}+K_{12}^{1}K_{11}^{2}+K_{12}%
^{2}K_{12}^{2}\\
&  =1.
\end{align*}

Choosing $K_{11}^{1}K_{11}^{1}=1$ we have $K_{12}^{2}K_{12}^{2}=$ $K_{12}%
^{1}K_{11}^{2}=K_{11}^{2}K_{12}^{1}=0$. This implies that the law of
composition of $S_{2,1,0}$ must satisfy%

\begin{align*}
s_{1}\diamondsuit s_{1}  &  =s_{1}\\
s_{1}\diamondsuit s_{2}  &  \neq s_{2}\longrightarrow s_{1}\diamondsuit
s_{2}=s_{1}%
\end{align*}

For the other double sum we have%

\begin{align*}
K_{2\gamma}^{\delta}K_{2\delta}^{\gamma}  &  =K_{21}^{\delta}K_{2\delta}%
^{1}+K_{22}^{\delta}K_{2\delta}^{2}\\
&  =K_{21}^{1}K_{21}^{1}+K_{21}^{2}K_{22}^{1}+K_{22}^{1}K_{21}^{2}+K_{22}%
^{2}K_{22}^{2}%
\end{align*}
which leads to%

\[
s_{2}\diamondsuit s_{2}=s_{1}%
\]
and for non diagonal double sum (con $K_{11}^{1}K_{11}^{1}=1$)%

\begin{align*}
K_{1\gamma}^{\delta}K_{2\delta}^{\gamma}  &  =K_{11}^{\delta}K_{2\delta}%
^{1}+K_{12}^{\delta}K_{2\delta}^{2}\\
&  =K_{11}^{1}K_{21}^{1}+K_{11}^{2}K_{22}^{1}+K_{12}^{1}K_{21}^{2}+K_{12}%
^{2}K_{22}^{2}\\
&  =K_{11}^{1}K_{21}^{1}+0+0+0\\
&  =1
\end{align*}

This leads to the semigroup%

\[
S_{2,1,0}=%
\begin{tabular}
[c]{ccc}%
$\diamondsuit$ & $\lambda_{1}$ & $\lambda_{2}$\\
$\lambda_{1}$ & $\lambda_{1}$ & $\lambda_{1}$\\
$\lambda_{2}$ & $\lambda_{1}$ & $\lambda_{1}$%
\end{tabular}
\text{ \ }%
\]

The $K_{12}^{2}K_{12}^{2}=1$ condition%

\begin{align*}
K_{1\gamma}^{\delta}K_{2\delta}^{\gamma}  &  =K_{11}^{\delta}K_{2\delta}%
^{1}+K_{22}^{\delta}K_{2\delta}^{2}\\
&  =K_{11}^{1}K_{21}^{1}+K_{11}^{2}K_{22}^{1}+K_{12}^{1}K_{21}^{2}+K_{12}%
^{2}K_{22}^{2}\\
&  =0+0+0+1\\
&  =1
\end{align*}
leads to the semigroup%

\[
S_{2,1,0}=%
\begin{tabular}
[c]{ccc}%
$\diamondsuit$ & $\lambda_{1}$ & $\lambda_{2}$\\
$\lambda_{1}$ & $\lambda_{2}$ & $\lambda_{2}$\\
$\lambda_{2}$ & $\lambda_{2}$ & $\lambda_{2}$%
\end{tabular}
\]

Finally to $H=0$ and $Q=0$ one has the condition%

\begin{align*}
\left(  K_{1\gamma}^{\delta}K_{1\delta}^{\gamma}\right)  \left(  K_{2\gamma
}^{\delta}K_{2\delta}^{\gamma}\right)   &  >\left(  K_{1\gamma}^{\delta
}K_{2\delta}^{\gamma}\right)  ^{2}\\
ac  &  >b^{2}%
\end{align*}
we use again $K_{i\gamma}^{\delta}K_{j\delta}^{\gamma}\in\left\{
0,1,2\right\}  $ to find semigroups satisfying this condition.%

\begin{align}
ac  &  >b^{2}\nonumber\\
2\cdot2  &  >1\nonumber\\
2\cdot1  &  >1\nonumber\\
1\cdot2  &  >1\nonumber\\
1\cdot1  &  >0\nonumber\\
2\cdot1  &  >0\nonumber\\
1\cdot2  &  >0\nonumber\\
2\cdot2  &  >0
\end{align}

This leads to the following matrices $M_{K}$%

\begin{align*}
\left(  K_{i\gamma}^{\delta}K_{j\delta}^{\gamma}\right)   &  =%
\begin{pmatrix}
2 & 1\\
1 & 2
\end{pmatrix}
\equiv M_{K1}\\
_{d}\left(  K_{i\gamma}^{\delta}K_{j\delta}^{\gamma}\right)   &  =%
\begin{pmatrix}
2 & 0\\
0 & 1
\end{pmatrix}
\\
&  \wedge\\
\left(  K_{i\gamma}^{\delta}K_{j\delta}^{\gamma}\right)   &  =%
\begin{pmatrix}
2 & 1\\
1 & 1
\end{pmatrix}
\equiv M_{K2}\\
_{d}\left(  K_{i\gamma}^{\delta}K_{j\delta}^{\gamma}\right)   &  =%
\begin{pmatrix}
\frac{\sqrt{5}}{2}+\frac{3}{2} & 0\\
0 & \frac{3}{2}-\frac{\sqrt{5}}{2}%
\end{pmatrix}
\\
&  \wedge\\
\left(  K_{i\gamma}^{\delta}K_{j\delta}^{\gamma}\right)   &  =%
\begin{pmatrix}
1 & 1\\
1 & 2
\end{pmatrix}
\equiv M_{K3}\\
_{d}\left(  K_{i\gamma}^{\delta}K_{j\delta}^{\gamma}\right)   &  =%
\begin{pmatrix}
\frac{3}{2}-\frac{\sqrt{5}}{2} & 0\\
0 & \frac{\sqrt{5}}{2}+\frac{3}{2}%
\end{pmatrix}
\\
&  \wedge\\
\left(  K_{i\gamma}^{\delta}K_{j\delta}^{\gamma}\right)   &  =%
\begin{pmatrix}
2 & 0\\
0 & 1
\end{pmatrix}
\equiv M_{K4}\\
&  \wedge\\
\left(  K_{i\gamma}^{\delta}K_{j\delta}^{\gamma}\right)   &  =%
\begin{pmatrix}
1 & 0\\
0 & 2
\end{pmatrix}
\equiv M_{K5}\\
&  \wedge\\
\left(  K_{i\gamma}^{\delta}K_{j\delta}^{\gamma}\right)   &  =%
\begin{pmatrix}
1 & 0\\
0 & 1
\end{pmatrix}
\equiv M_{K6}\\
&  \wedge\\
\left(  K_{i\gamma}^{\delta}K_{j\delta}^{\gamma}\right)   &  =%
\begin{pmatrix}
2 & 0\\
0 & 2
\end{pmatrix}
\equiv M_{K7}%
\end{align*}

To check whether there is any semigroup of order two that leads us to these
matrices $M_{K}$ we follow a similar process to the one above case, with
$K_{i\gamma}^{\delta}K_{j\delta}^{\gamma}$. If $K_{1\gamma}^{\delta}%
K_{1\delta}^{\gamma}=a=2$ we have four possibilities%

\begin{align*}
ac  &  >b^{2}\\
2\cdot2  &  >1\\
2\cdot1  &  >1\\
1\cdot2  &  >1\\
2\cdot1  &  >0\\
1\cdot2  &  >0\\
2\cdot2  &  >0
\end{align*}
corresponding to $K_{11}^{1}K_{11}^{1}\neq0$ $\wedge$ $K_{12}^{2}K_{12}%
^{2}\neq0$ or $K_{11}^{2}K_{12}^{1}\neq0$%

\begin{align*}
K_{1\gamma}^{\delta}K_{1\delta}^{\gamma}  &  =K_{11}^{\delta}K_{1\delta}%
^{1}+K_{12}^{\delta}K_{1\delta}^{2}\\
&  =K_{11}^{1}K_{11}^{1}+K_{11}^{2}K_{12}^{1}\\
&  +K_{12}^{1}K_{11}^{2}+K_{12}^{2}K_{12}^{2}\\
&  =2
\end{align*}

Choosing%

\begin{align*}
K_{11}^{2}K_{12}^{1}  &  \neq0\\
K_{11}^{2}  &  \neq0\text{ \ }\wedge\text{\ }K_{12}^{1}\neq0
\end{align*}
we find%

\begin{align*}
K_{2\gamma}^{\delta}K_{2\delta}^{\gamma}  &  =K_{21}^{\delta}K_{2\delta}%
^{1}+K_{22}^{\delta}K_{2\delta}^{2}\\
&  =K_{21}^{1}K_{21}^{1}+K_{21}^{2}K_{22}^{1}\\
&  +K_{22}^{1}K_{21}^{2}+K_{22}^{2}K_{22}^{2}=\left\{  1,2\right\}
\end{align*}

\[
K_{22}^{2}K_{22}^{2}\in\left\{  0,1\right\}  .
\]
To $K_{22}^{2}\neq0$ , $K_{2\gamma}^{\delta}K_{2\delta}^{\gamma}=2$ we have%

\begin{align*}
K_{1\gamma}^{\delta}K_{2\delta}^{\gamma}  &  =K_{11}^{\delta}K_{2\delta}%
^{1}+K_{12}^{\delta}K_{2\delta}^{2}\\
&  =K_{11}^{1}K_{21}^{1}+K_{11}^{2}K_{22}^{1}\\
&  +K_{12}^{1}K_{21}^{2}+K_{12}^{2}K_{22}^{2}\\
&  =0
\end{align*}

\begin{align*}
K_{11}^{1}K_{21}^{1}  &  =0\\
K_{12}^{1}K_{21}^{2}  &  =0\\
K_{11}^{2}K_{22}^{1}  &  =0\\
K_{12}^{2}K_{22}^{2}  &  =0
\end{align*}
So that%

\begin{equation}
S_{2,0,0}=%
\begin{tabular}
[c]{||c||c|c|}\hline\hline
$\diamondsuit$ & $\lambda_{1}$ & \multicolumn{1}{||c||}{$\lambda_{2}$%
}\\\hline\hline
$\lambda_{1}$ & $\lambda_{2}$ & $\lambda_{1}$\\\hline
$\lambda_{2}$ & $\lambda_{1}$ & $\lambda_{2}$\\\hline
\end{tabular}
\longrightarrow M_{K}=%
\begin{pmatrix}
2 & 0\\
0 & 2
\end{pmatrix}
.
\end{equation}
For choice $K_{22}^{2}=0$ we have $K_{1\gamma}^{\delta}K_{2\delta}^{\gamma}$
is non-zero, because $K_{11}^{2}K_{22}^{1}=1$. With this choice, the set
$K\neq0\longrightarrow\left\{  K_{12}^{1},~K_{11}^{2},~K_{21}^{1},~K_{22}%
^{1}\right\}  $ is obtained. The matrix $M_{K}$ and semigroup in this case are%

\begin{equation}
S_{2,0,0}=%
\begin{tabular}
[c]{||c||c|c|}\hline\hline
$\diamondsuit$ & $\lambda_{1}$ & \multicolumn{1}{||c||}{$\lambda_{2}$%
}\\\hline\hline
$\lambda_{1}$ & $\lambda_{2}$ & $\lambda_{1}$\\\hline
$\lambda_{2}$ & $\lambda_{1}$ & $\lambda_{1}$\\\hline
\end{tabular}
\longrightarrow M_{K}=%
\begin{pmatrix}
2 & 1\\
1 & 1
\end{pmatrix}
\end{equation}
By choosing now, since the beginning $K_{11}^{1}K_{11}^{1}\neq0$ and
$K_{12}^{2}K_{12}^{2}\neq0$, we have%

\begin{align*}
K_{2\gamma}^{\delta}K_{2\delta}^{\gamma}  &  =K_{21}^{\delta}K_{2\delta}%
^{1}+K_{22}^{\delta}K_{2\delta}^{2}\\
&  =K_{21}^{1}K_{21}^{1}+K_{21}^{2}K_{22}^{1}+K_{22}^{1}K_{21}^{2}+K_{22}%
^{2}K_{22}^{2}\\
&  =\left\{  1,2\right\}
\end{align*}
is equal to two for $K_{22}^{1}\neq0$ and is equal to one for $K_{22}^{1}=0$
and $K_{22}^{2}\neq0$. Using $K_{22}^{1}\neq0,$ $K_{2\gamma}^{\delta
}K_{2\delta}^{\gamma}=1$ we find%

\begin{align*}
K_{1\gamma}^{\delta}K_{2\delta}^{\gamma}  &  =K_{11}^{\delta}K_{2\delta}%
^{1}+K_{12}^{\delta}K_{2\delta}^{2}\\
&  =K_{11}^{1}K_{21}^{1}+K_{11}^{2}K_{22}^{1}+K_{12}^{1}K_{21}^{2}+K_{12}%
^{2}K_{22}^{2}\\
&  =0
\end{align*}

So that:%

\begin{equation}
S_{2,0,0}=%
\begin{tabular}
[c]{||c||c|c|}\hline\hline
$\diamondsuit$ & $\lambda_{1}$ & \multicolumn{1}{||c||}{$\lambda_{2}$%
}\\\hline\hline
$\lambda_{1}$ & $\lambda_{1}$ & $\lambda_{2}$\\\hline
$\lambda_{2}$ & $\lambda_{2}$ & $\lambda_{1}$\\\hline
\end{tabular}
\longrightarrow M_{K}=%
\begin{pmatrix}
2 & 0\\
0 & 2
\end{pmatrix}
\end{equation}

Following the same procedure, we find the remaining possible semigroups. In
summary, we have that the possible semigroups are%

\begin{align}
A  &  =%
\begin{tabular}
[c]{||c||c|c|}\hline\hline
$\diamondsuit$ & $\lambda_{1}$ & \multicolumn{1}{||c||}{$\lambda_{2}$%
}\\\hline\hline
$\lambda_{1}$ & $\lambda_{2}$ & $\lambda_{1}$\\\hline
$\lambda_{2}$ & $\lambda_{1}$ & $\lambda_{2}$\\\hline
\end{tabular}
\longrightarrow M_{K_{A}}=%
\begin{pmatrix}
2 & 0\\
0 & 2
\end{pmatrix}
=\tilde{M}_{K_{A}}\longleftarrow%
\begin{tabular}
[c]{||c||c|c|}\hline\hline
$\diamondsuit$ & $\lambda_{1}$ & \multicolumn{1}{||c||}{$\lambda_{2}$%
}\\\hline\hline
$\lambda_{1}$ & $\lambda_{1}$ & $\lambda_{2}$\\\hline
$\lambda_{2}$ & $\lambda_{2}$ & $\lambda_{1}$\\\hline
\end{tabular}
\nonumber\\
B  &  =%
\begin{tabular}
[c]{||c||c|c|}\hline\hline
$\diamondsuit$ & $\lambda_{1}$ & \multicolumn{1}{||c||}{$\lambda_{2}$%
}\\\hline\hline
$\lambda_{1}$ & $\lambda_{2}$ & $\lambda_{1}$\\\hline
$\lambda_{2}$ & $\lambda_{1}$ & $\lambda_{1}$\\\hline
\end{tabular}
\longrightarrow M_{K_{B}}=%
\begin{pmatrix}
2 & 1\\
1 & 1
\end{pmatrix}
=\tilde{M}_{K_{B}}\longleftarrow%
\begin{tabular}
[c]{||c||c|c|}\hline\hline
$\diamondsuit$ & $\lambda_{1}$ & \multicolumn{1}{||c||}{$\lambda_{2}$%
}\\\hline\hline
$\lambda_{1}$ & $\lambda_{1}$ & $\lambda_{2}$\\\hline
$\lambda_{2}$ & $\lambda_{2}$ & $\lambda_{2}$\\\hline
\end{tabular}
\nonumber\\
C  &  =%
\begin{tabular}
[c]{||c||c|c|}\hline\hline
$\diamondsuit$ & $\lambda_{1}$ & \multicolumn{1}{||c||}{$\lambda_{2}$%
}\\\hline\hline
$\lambda_{1}$ & $\lambda_{1}$ & $\lambda_{1}$\\\hline
$\lambda_{2}$ & $\lambda_{1}$ & $\lambda_{2}$\\\hline
\end{tabular}
\longrightarrow M_{K_{C}}=%
\begin{pmatrix}
1 & 1\\
1 & 2
\end{pmatrix}
\nonumber\\
D  &  =%
\begin{tabular}
[c]{||c||c|c|}\hline\hline
$\diamondsuit$ & $\lambda_{1}$ & \multicolumn{1}{||c||}{$\lambda_{2}$%
}\\\hline\hline
$\lambda_{1}$ & $\lambda_{2}$ & $\lambda_{2}$\\\hline
$\lambda_{2}$ & $\lambda_{2}$ & $\lambda_{1}$\\\hline
\end{tabular}
\longrightarrow M_{K_{D}}=%
\begin{pmatrix}
1 & 1\\
1 & 2
\end{pmatrix}
\nonumber
\end{align}

\begin{equation}
\left(  g_{AB}\right)  _{S_{2,0,0}\otimes~so\left(  3\right)  }=-1\cdot%
\begin{pmatrix}
2 & 1 & 0 & 0 & 0 & 0\\
1 & 1 & 0 & 0 & 0 & 0\\
0 & 0 & 2 & 1 & 0 & 0\\
0 & 0 & 1 & 1 & 0 & 0\\
0 & 0 & 0 & 0 & 2 & 1\\
0 & 0 & 0 & 0 & 1 & 1
\end{pmatrix}
\end{equation}

Where, for example, $A$ to coincides with the cyclic group $%
\mathbb{Z}
_{2}$%

\begin{equation}
S_{2,0,0}\equiv A=%
\begin{tabular}
[c]{||c||c|c|}\hline\hline
$\diamondsuit$ & $\lambda_{1}$ & \multicolumn{1}{||c||}{$\lambda_{2}$%
}\\\hline\hline
$\lambda_{1}$ & $\lambda_{1}$ & $\lambda_{2}$\\\hline
$\lambda_{2}$ & $\lambda_{2}$ & $\lambda_{1}$\\\hline
\end{tabular}
\longrightarrow%
\begin{tabular}
[c]{||c||c|c|}\hline\hline
$\diamondsuit$ & $\lambda_{0}$ & \multicolumn{1}{||c||}{$\lambda_{1}$%
}\\\hline\hline
$\lambda_{0}$ & $\lambda_{0}$ & $\lambda_{1}$\\\hline
$\lambda_{0}$ & $\lambda_{1}$ & $\lambda_{0}$\\\hline
\end{tabular}
=%
\mathbb{Z}
_{2}~.
\end{equation}

The natural question is: multiplication tables $B,~C,~D$ are semigroups
multiplication tables?. To answer let us see if these multiplication satisfy
associativity. It is direct to verify that multiplication tables $B$ and $D$
do not satisfy the associative property. So that the associated semigroup
table $C$ lead to a Lie algebra, and semisimple compact, like $%
\mathbb{Z}
_{2}$ semigroup. The question now is whether the algebras obtained by
expansion using $%
\mathbb{Z}
_{2}$ and $C$ are isomorphic or not. The answer can be found in two ways. The
first is to use the character table shown above. From this table we can verify
by inspection that $\chi=-~6$ univocally characterizes $so\left(  4\right)  $
and its isomorphic forms.

The other way to tell if the two Lie algebras are isomorphic or not, is using
$MAGMA$ \cite{magm}. With this program you can check if the product spaces,
are or are not Lie algebras and thus check whether the tables obtained
correspond to semigroups.

The above outcomes allow to state: "The semigroup which leads from a Lie
algebra to another by S-expansion method is not necessarily unique."

Following the procedure described above is possible to obtain the possible
S-expansions that could lead from $so\left(  n\right)  $ to $so\left(
n+l\right)  $.%

\begin{equation}%
\begin{tabular}
[c]{||l||l|l|l|l||}\hline\hline
$so\left(  n\right)  $ & $so\left(  n+l\right)  $ & \multicolumn{1}{||l|}{$P$}
& \multicolumn{1}{||l|}{$H$} & \multicolumn{1}{||l||}{$Q$}\\\hline\hline
$\mathbf{3}$ & $\mathbf{4}$ & $\mathbf{2}$ & $\mathbf{0}$ & $\mathbf{0}%
$\\\hline
$3$ & $6$ & $5$ & $0$ & $0$\\\hline
$3$ & $7$ & $7$ & $0$ & $0$\\\hline
$3$ & $9$ & $12$ & $0$ & $0$\\\hline
$3$ & $10$ & $15$ & $0$ & $0$\\\hline
$3$ & $12$ & $22$ & $0$ & $0$\\\hline
$4$ & $9$ & $6$ & $0$ & $0$\\\hline
$4$ & $12$ & $11$ & $0$ & $0$\\\hline
$4$ & $13$ & $13$ & $0$ & $0$\\\hline
$4$ & $16$ & $20$ & $0$ & $0$\\\hline
$5$ & $16$ & $12$ & $0$ & $0$\\\hline
$5$ & $20$ & $19$ & $0$ & $0$\\\hline
$5$ & $21$ & $21$ & $0$ & $0$\\\hline
$6$ & $10$ & $3$ & $0$ & $0$\\\hline
$6$ & $15$ & $7$ & $0$ & $0$\\\hline
$6$ & $16$ & $8$ & $0$ & $0$\\\hline\hline
\end{tabular}
\tag{$Tabla~I$}%
\end{equation}

\section{\textbf{No-simplicity of S-expanded algebra}}

From the above table we can see that there are several possible S-expansions
that we could perform. However there are some restrictions that must be
satisfied. A simple Lie algebra is one that has no nontrivial ideals and that
can not be expressed as a direct sum of other Lie algebras. This fact leads to
the important result that an Lie algebra obtained by $S$-expansion of another
Lie algebra can not be simple.

A square matrix $\left(  M_{f}\right)  _{S}$ of order $P$ and range $P$, will
have $P$ nonzero eigenvalues and can be expressed in the form%

\begin{equation}
\left(  M_{f}\right)  _{S,d}=%
\begin{pmatrix}
\lambda_{0} &  & O\\
& \ddots & \\
O &  & \lambda_{P-1}%
\end{pmatrix}
_{P\times P}. \label{one}%
\end{equation}
The Kronecker product between $\left(  M_{f}\right)  _{S,d}$ and an arbitrary
order matrix $R$ such as $Ad\left(  \mathcal{G}\right)  $ is given by%

\begin{align}%
\begin{pmatrix}
\lambda_{0} &  & O\\
& \ddots & \\
O &  & \lambda_{P-1}%
\end{pmatrix}
_{P\times P}\otimes Ad\left(  \mathcal{G}\right)  ~~  &  =~~%
\begin{pmatrix}
\lambda_{0}\left(  \bar{M}\right)  &  & O\\
& \ddots & \\
O &  & \lambda_{P-1}\left(  \bar{M}\right)
\end{pmatrix}
_{PR\times PR}\nonumber\\
&  =~\underset{P-times}{\underbrace{\lambda_{0}\left(  \bar{M}\right)
\oplus\cdots\oplus\lambda_{P-1}\left(  \bar{M}\right)  }}~\equiv Ad\left(
\mathcal{\bar{G}}\right)  \label{two}%
\end{align}

This means that if $\left(  \bar{M}\right)  $ is the adjoint representation of
an arbitrary Lie algebra $\mathcal{G}$ and if $\left(  M_{f}\right)  _{S}$ is
a faithful matrix representation of an abelian, discrete and finite semigroup
$S$, then $Ad\left(  \mathcal{\bar{G}}\right)  $ is the adjoint representation
of a non-simple Lie algebra, given by the direct sum of $P$ Lie algebras
$\mathcal{G}$ (which can be simple or not). This will occur when the rank of
the matrix $\left(  M_{f}\right)  _{S}$ is equal to the number of elements of
the semigroup $S$.

This result leads to state the following

\textbf{Theorem:} If S is a finite, discrete and abelian semigroup and if
$\mathcal{G}$ is an arbitrary Lie algebra, then the product space
$S\otimes\mathcal{G}$ is a non-simple Lie algebra consisting of the direct sum
of $P$ original Lie algebras $\mathcal{G}$, where $P$ is the number of
elements of the semigroup $S$.

\textbf{Proof:} The faithful matrix representation of any abelian, discrete
and finite semigroup has the form%

\begin{equation}
\left(  M_{f}\right)  _{S}=%
\begin{pmatrix}
\left(  K_{r}\right)  _{0}^{0} & \cdots & \left(  K_{u}\right)  _{0}^{x} &
\cdots & \left(  K_{x}\right)  _{0}^{P-1}\\
\vdots & \ddots &  &  & \vdots\\
\left(  K_{s}\right)  _{i}^{0} & \cdots & \left(  K_{v}\right)  _{i}^{x} &  &
\left(  K_{y}\right)  _{i}^{P-1}\\
\vdots &  &  & \ddots & \vdots\\
\left(  K_{t}\right)  _{p-1}^{0} & \cdots & \left(  K_{w}\right)  _{i}^{x} &
\cdots & \left(  K_{z}\right)  _{P-1}^{P-1}%
\end{pmatrix}
_{P\times P}. \label{three}%
\end{equation}

If the matrix has a lower rank than $P$ then will have $(i)$ two equal or
proportional rows or alternatively $(ii)$ a third row that is a linear
combination of other linearly independent rows.

$\mathbf{(i)}$ \ \textbf{Two equal or proportional rows}: We will use the
indices $i,j,k,r$ with $j\not =r$ (e.g. $j<r$). The operation between the
$i$-$th$ and $j$-$th$ element of the semigroup results in the $k$-$th$ element
of the semigroup. The corresponding matrix element is $\left(  K_{i}\right)
_{j}^{k}$, which is located in the $j$-$th$ row and $k$-$th$ column of the
matrix $\left(  M_{f}\right)  _{S}$. \ \ If there is an $i$-$th$ row equal or
proportional to it, then there is also an element of the form $C\left(
K_{j}\right)  _{i}^{r}$, which is not in the$k$-$th$ column because if the
rows were equal, then the element in the row belonging to the $k$-$th$ column
will have the form $\left(  K_{i}\right)  _{i}^{k}$, for which $k\not =r$.
This has the consequence that $\lambda_{i}\diamondsuit\lambda_{j}=\lambda_{k}$
and $\lambda_{j}\diamondsuit\lambda_{i}=\lambda_{r}$ with $\lambda_{k}%
\not =\lambda_{r}$ implying that $\lambda_{i}\diamondsuit\lambda_{j}%
\not =\lambda_{j}\diamondsuit\lambda_{i}$.

This contradicts the condition of abelian semigroup $S$. So the matrix
$\left(  M_{f}\right)  _{S}$ has no equal (or proportional) rows. Similarly it
is proved that the matrix $\left(  M_{f}\right)  _{S}$ does not have equal (or
proportional) columns.

\textbf{(b) A row is a linear combination of linearly independent rows:} Note
that the $j$-$th$ element of the $i$-$th$ row has the form $\left(
K_{j}\right)  _{i}^{y}$. This means that if another element in the same row
has the form $\left(  K_{j}\right)  _{i}^{z}$ to different columns, i.e., for
$y\not =z$, then this implies $\lambda_{j}\diamondsuit\lambda_{i}=\lambda_{y}$
y $\lambda_{j}\diamondsuit\lambda_{i}=\lambda_{z}$. \ \ But $\lambda_{y}%
\not =\lambda_{z}$. This has the consequence that $\lambda_{j}\diamondsuit
\lambda_{i}\not =\lambda_{j}\diamondsuit\lambda_{i}$, which is absurd.
Additionally this would imply that the internal binary operation
$\diamondsuit$ semigroup is not univocally defined for each pair of elements.
This allows us to ensure that we never have two or more $\left(  K_{j}\right)
_{i}^{y}$ -selectores associated with the same binary operation $\lambda
_{j}\diamondsuit\lambda_{i}$ in a row.

Thus we have the row that is generated by the linear combination of at least
two independent rows in the matrix $\left(  M_{f}\right)  _{S}$ will always
have an element of the form%

\begin{equation}
C^{i_{0}}\left(  K_{q}\right)  _{i_{0}}^{y}+\cdots+C^{i_{P-1}}\left(
K_{x}\right)  _{i_{P-1}}^{y}%
\end{equation}
and another element of the form%

\begin{equation}
C^{i_{0}}\left(  K_{q}\right)  _{i_{0}}^{z}+\cdots+C^{i_{P-1}}\left(
K_{x}\right)  _{i_{P-1}}^{z}.
\end{equation}

So that for the elements of the semigroup one has%

\begin{align}
\lambda_{q}\diamondsuit\lambda_{i_{0}}  &  =\lambda_{y}\text{ \ \ }%
\wedge\text{ \ \ }\lambda_{q}\diamondsuit\lambda_{i_{0}}=\lambda
_{z}\nonumber\\
&  \vdots\\
\lambda_{x}\diamondsuit\lambda_{i_{P-1}}  &  =\lambda_{y}\text{ \ \ }%
\wedge\text{ \ \ }\lambda_{x}\diamondsuit\lambda_{i_{P-1}}=\lambda
_{z},\nonumber
\end{align}
i.e.,%

\begin{align}
\lambda_{q}\diamondsuit\lambda_{i_{0}}  &  \not =\lambda_{q}\diamondsuit
\lambda_{i_{0}}\nonumber\\
&  \vdots\\
\lambda_{x}\diamondsuit\lambda_{i_{P-1}}  &  \not =\lambda_{x}\diamondsuit
\lambda_{i_{P-1}}.\nonumber
\end{align}

This leads to the absurd result that there are more than one internal binary
operations that are not univocally defined.

These results allow us to affirm that the faithful matrix representation
$\left(  M_{f}\right)  _{S}$ of a semigroup of $P$ elements will always be of
rank $P$. Therefore, the diagonal form will have $P$ nonzero eigenvalues and
the product space $S\otimes\mathcal{G}$ will be the direct sum of $P$ times
the original Lie algebra $\mathcal{G}.$ This means that $S\otimes\mathcal{G} $
\ will be a non-simple Lie algebra.

\section{\textbf{Concluding Remarks}}

In this work we have reviewed some concepts of the theory of Lie algebras and
the main aspects of the S-expansion procedure. Probably the most important
result of this article is the fact that the S-expansion procedure affects the
geometry of a Lie group: was found how changing the magnitude of a vector and
the angle between two vectors. \ Was outlined, via an example, a method for
determining the semigroup, which would provide a Lie algebra from
another\textbf{\ }and then\ proved that a Lie algebra obtained from another
Lie algebra via S-expansion is a non-simple Lie algebra.

A future work could be consider the geom\'{e}trical analysis of the $0_{S}%
$-Resonant procedure and also get the selection rules to determine when it%
\'{}%
s possible to get another algebra and then obtain the suitable semigroups and
the partitions that are necessary to obtain it finally in each case (working progress).

This work was supported in part by FONDECYT Grants N$^{0}$ 1130653. Two of the
authors (MC, DMP) were supported by grants from the Comisi\'{o}n Nacional de
Investigaci\'{o}n Cient\'{\i}fica y Tecnol\'{o}gica CONICYT and from the
Universidad de Concepci\'{o}n, Chile.

\section{Appendix A: Inner product in the\ S-expanded Lie algebra}

Let $\left\{  \lambda_{\alpha}\right\}  $ be an abelian semigroup with
two-selectors $K_{\alpha\beta}^{\gamma}$ and $\mathfrak{g}$ a Lie algebra
withbasis $\left\{  T_{A}\right\}  $ and structure constants $C_{AB}^{\text{
\ \ }C}.$ Denote a basis elements of the $S$-expanded Lie algebra
$S\otimes\mathfrak{g}$ by $T_{\left(  a,\alpha\right)  \text{ }}\equiv
\lambda_{\alpha}T_{a}.$ \emph{\ }The\emph{\ }inner product between the
$X=v^{\left(  \alpha,a\right)  }T_{\left(  a,\alpha\right)  \text{ }}$vectors
of the $S\otimes\mathfrak{g}$\ space is given by%

\begin{equation}
\left(  X,X\right)  _{S-\exp}\equiv tr\left(  R(X)R(X)\right)  =v^{\left(
\alpha,a\right)  }v^{\left(  \beta,b\right)  }K_{\alpha\gamma}^{\delta
}K_{\beta\delta}^{\gamma}~\left(  X_{a},X_{b}\right)  \label{1}%
\end{equation}
where $v^{\left(  \alpha,a\right)  }$ are the $S\otimes\mathfrak{g}$
coordinates, $K_{\beta\delta}^{\gamma}$, the $K$-selectors defined by product
between the semigroups elements and $\left(  X_{a},X_{b}\right)  $ is the
Killing-Cartan inner product defined in the Lie algebra $\mathfrak{g}$. Now we
will show that the product (\ref{1}) satisfy the axioms of the inner product.
In fact,

$(a)$ $\ \ If$ $X,Y,Z\in S\otimes\mathfrak{g,}$ then $\left(  X+Y,Z\right)
=\left(  X,Z\right)  +\left(  Y,Z\right)  $

\textbf{Proof: }Since,%

\begin{align*}
\left(  X+Y,Z\right)  _{S-\exp}  &  =tr\left(  R\left(  X+Y\right)  R\left(
Z\right)  \right) \\
&  =tr\left(  R\left(  X\right)  R\left(  Z\right)  +R\left(  Y\right)
R\left(  Z\right)  \right)
\end{align*}
we have%

\begin{align*}
\left(  X+Y,Z\right)  _{S-\exp}  &  =tr\left(  v^{\left(  \alpha,a\right)
}R\left(  X_{\left(  \alpha,a\right)  }\right)  v^{\left(  \gamma,c\right)
}R\left(  Z_{\left(  \gamma,c\right)  }\right)  \right) \\
&  +tr\left(  v^{\left(  \beta,b\right)  }R\left(  Y_{\left(  \beta,b\right)
}\right)  v^{\left(  \gamma,c\right)  }R\left(  Z_{\left(  \gamma,c\right)
}\right)  \right) \\
&  =v^{\left(  \alpha,a\right)  }v^{\left(  \gamma,c\right)  }K_{\alpha\delta
}^{\varepsilon}K_{\gamma\varepsilon}^{\delta}\left(  C_{a}\right)  _{d}%
^{e}\left(  C_{c}\right)  _{e}^{d}\\
&  +v^{\left(  \beta,b\right)  }v^{\left(  \gamma,c\right)  }K_{\beta\delta
}^{\varepsilon}K_{\gamma\varepsilon}^{\delta}\left(  C_{b}\right)  _{d}%
^{e}\left(  C_{c}\right)  _{e}^{d}\\
&  =v^{\left(  \alpha,a\right)  }v^{\left(  \gamma,c\right)  }K_{\alpha\delta
}^{\varepsilon}K_{\gamma\varepsilon}^{\delta}~tr\left(  R\left(  X_{a}\right)
R\left(  Z_{c}\right)  \right) \\
&  +v^{\left(  \beta,b\right)  }v^{\left(  \gamma,c\right)  }K_{\beta\delta
}^{\varepsilon}K_{\gamma\varepsilon}^{\delta}~tr\left(  R\left(  Y_{b}\right)
R\left(  Z_{c}\right)  \right) \\
&  =v^{\left(  \alpha,a\right)  }v^{\left(  \gamma,c\right)  }K_{\alpha\delta
}^{\varepsilon}K_{\gamma\varepsilon}^{\delta}~\left(  X_{a},Z_{c}\right) \\
&  +v^{\left(  \beta,b\right)  }v^{\left(  \gamma,c\right)  }K_{\beta\delta
}^{\varepsilon}K_{\gamma\varepsilon}^{\delta}~\left(  Y_{b},Z_{c}\right) \\
&  =\left(  X,Z\right)  _{S-\exp}+\left(  Y,Z\right)  _{S-\exp}\text{ ,}%
\end{align*}
where we have used $(i)$\emph{\ }linearity of the adjoint representation,
$(ii)$ linearity of the trace, $(iii)$ the definition of the Killing-Cartan
inner product in the algebra $\mathfrak{g}.$

$(b)$ If $X,Y\in S\otimes\mathfrak{g,}$ then $\left(  \alpha X,Y\right)
=\alpha\left(  X,Y\right)  $

Proof:%

\begin{align*}
\left(  \alpha X,Y\right)  _{S-\exp}  &  =tr\left(  v^{\left(  \alpha
,a\right)  }R\left(  \alpha X_{\left(  \alpha,a\right)  }\right)  v^{\left(
\beta,b\right)  }R\left(  Y_{\left(  \beta,b\right)  }\right)  \right) \\
&  =\alpha v^{\left(  \alpha,a\right)  }v^{\left(  \beta,b\right)  }%
K_{\alpha\gamma}^{\delta}K_{\beta\delta}^{\gamma}\left(  C_{a}\right)
_{c}^{d}\left(  C_{b}\right)  _{d}^{c}\\
&  =\alpha v^{\left(  \alpha,a\right)  }v^{\left(  \beta,b\right)  }%
K_{\alpha\gamma}^{\delta}K_{\beta\delta}^{\gamma}~\left(  X_{a},X_{b}\right)
\\
&  =\alpha\left(  X,Y\right)  _{S-\exp}\text{ \ ,}%
\end{align*}

where we have used $(i)$ property of the trace, $(ii)$ the definition of the
Killing-Cartan inner product in the algebra $\mathfrak{g}$.

$(c)$ If $X,Y\in S\otimes\mathfrak{g,}$ then $\left(  X,Y\right)  =\left(
Y,X\right)  $

Proof:%

\begin{align*}
\left(  X,Y\right)  _{S-\exp}=tr\left(  R\left(  X\right)  R\left(  Y\right)
\right)   &  =tr\left(  v^{\left(  \alpha,a\right)  }R\left(  X_{\left(
\alpha,a\right)  }\right)  v^{\left(  \beta,b\right)  }R\left(  Y_{\left(
\beta,b\right)  }\right)  \right) \\
&  =v^{\left(  \alpha,a\right)  }v^{\left(  \beta,b\right)  }K_{\alpha\gamma
}^{\delta}\left(  C_{a}\right)  _{c}^{d}K_{\beta\delta}^{\gamma}\left(
C_{b}\right)  _{d}^{c}\\
&  =v^{\left(  \alpha,a\right)  }v^{\left(  \beta,b\right)  }K_{\alpha\gamma
}^{\delta}K_{\beta\delta}^{\gamma}~tr\left(  R\left(  X_{a}\right)  R\left(
Y_{b}\right)  \right) \\
&  =v^{\left(  \alpha,a\right)  }v^{\left(  \beta,b\right)  }K_{\alpha\gamma
}^{\delta}K_{\beta\delta}^{\gamma}~tr\left(  R\left(  Y_{b}\right)  R\left(
X_{a}\right)  \right) \\
&  =v^{\left(  \beta,b\right)  }v^{\left(  \alpha,a\right)  }K_{\beta\delta
}^{\gamma}K_{\alpha\gamma}^{\delta}~\left(  X_{b},X_{a}\right) \\
&  =\left(  Y,X\right)  _{S-\exp}\text{ ,}%
\end{align*}

where we have used the fact that $(i)$ $R\left(  X_{a}\right)  $%
\emph{\ }and\emph{\ }$R\left(  Y_{b}\right)  $\emph{\ }are square matrices,
$(ii)$ $\left(  K_{\beta}\right)  _{\delta}^{\gamma}$\emph{\ }are square
matrices and forms a faithful representation of the semigroup elements, so
that\emph{\ }$K_{\alpha\gamma}^{\delta}K_{\beta\delta}^{\gamma}~=K_{\beta
\delta}^{\gamma}K_{\alpha\gamma}^{\delta}.$

\section{Appendix B: Invariance of the product $\left(  X,Y\right)  _{S-\exp}%
$}

The scalar product of arbitrary two elements $X=X^{a}T_{a}$ and $Y=Y^{b}T_{b}
$ of the finite-dimensional Lie algebra $g$ is given by the Killing form
\begin{equation}
(X,Y):=Tr(adX,adY)
\end{equation}

We know that scalar product is invariant under the action of the Lie group $G
$, if
\begin{equation}
\left(  gXg^{-1},gYg^{-1}\right)  =\left(  X,Y\right)
\end{equation}
where $g$ is an element of the $G$ group. \ For infinitesimal case it is
equivalent to
\begin{equation}
([X,Y],Z)=(X,[Y,Z]).
\end{equation}

We will use this condition to show that the Killing-Cartan product $\left(
X,X\right)  _{S-\exp}$ is invariant under the action of the Lie group $S\times
G$ , that is we will use $([X,Y],Z)_{S-\exp}=(X,[Y,Z])_{S-\exp}$ . This means
that we must show that%
\begin{equation}
Tr(ad[X_{\left(  \alpha,A\right)  },X_{\left(  \beta,B\right)  }]\cdot
adX_{\left(  \gamma,C\right)  })=Tr(adX_{\left(  \alpha,A\right)  }\cdot
ad[X_{\left(  \beta,B\right)  },X_{\left(  \gamma,C\right)  }])
\end{equation}
where $X_{\left(  \alpha,A\right)  }=\lambda_{\alpha}X_{A}$. \ 

Since,
\begin{equation}
\lbrack X_{\left(  \alpha,A\right)  },X_{\left(  \beta,B\right)  }]=C_{\left(
\alpha,A\right)  \left(  \beta,B\right)  }^{\left(  \gamma,C\right)
}X_{\left(  \gamma,C\right)  },
\end{equation}
we have%
\begin{equation}
ad[X_{\left(  \alpha,A\right)  },X_{\left(  \beta,B\right)  }]=C_{\left(
\alpha,A\right)  \left(  \beta,B\right)  }^{\left(  \gamma,C\right)
}adX_{\left(  \gamma,C\right)  },
\end{equation}
so that%
\begin{equation}
C_{\left(  \alpha,A\right)  \left(  \beta,B\right)  }^{\left(  \delta
,D\right)  }Tr\left[  (ad\left(  X_{\left(  \delta,D\right)  }\right)  \cdot
ad\left(  X_{\left(  \gamma,C\right)  }\right)  )\right]  =C_{\left(
\beta,B\right)  \left(  \gamma,C\right)  }^{\left(  \delta,D\right)
}Tr\left[  (adX_{\left(  \alpha,A\right)  }\cdot adX_{\left(  \delta,D\right)
})\right]  .
\end{equation}
Taking into account that
\begin{equation}
ad\left(  X_{\left(  \gamma,C\right)  }\right)  =C_{\left(  \gamma,C\right)
\left(  \rho,L\right)  }^{\left(  \lambda,J\right)  },
\end{equation}
we can see that%

\begin{equation}
C_{\left(  \alpha,A\right)  \left(  \beta,B\right)  }^{\text{
\ \ \ \ \ \ \ \ \ \ }\left(  \delta,D\right)  }C_{\left(  \delta,D\right)
\left(  \epsilon,E\right)  }^{\text{ \ \ \ \ \ \ \ \ }\left(  \varepsilon
,F\right)  }C_{\left(  \gamma,C\right)  \left(  \varepsilon,F\right)
}^{\text{ \ \ \ \ \ \ \ \ }\left(  \epsilon,E\right)  }=C_{\left(
\beta,B\right)  \left(  \gamma,C\right)  }^{\text{ \ \ \ \ \ \ \ \ \ }\left(
\delta,D\right)  }C_{\left(  \alpha,A\right)  \left(  \epsilon,E\right)
}^{\text{ \ \ \ \ \ \ \ \ }\left(  \varepsilon,F\right)  }C_{\left(
\delta,D\right)  \left(  \varepsilon,F\right)  }^{\text{ \ \ \ \ \ \ \ }%
\left(  \epsilon,E\right)  }.
\end{equation}

Since $C_{\left(  \alpha,A\right)  \left(  \beta,B\right)  }^{.........\left(
\delta,D\right)  }=K_{\alpha\beta}^{\delta}C_{AB}^{..D}$ , we can write%
\[
K_{\alpha\beta}^{\delta}K_{\delta\epsilon}^{\varepsilon}K_{\gamma\varepsilon
}^{\epsilon}C_{AB}^{..D}C_{DE}^{..F}C_{CF}^{..E}=K_{\beta\gamma}^{\delta
}K_{\alpha\epsilon}^{\varepsilon}K_{\delta\varepsilon}^{\epsilon}C_{BC}%
^{..D}C_{AE}^{..F}C_{DF}^{..E}.
\]

But since the Killing-Cartan product in $\mathfrak{g}$ is invariant under
transformations in $\mathfrak{g,}$ i.e.,%

\begin{equation}
C_{AB}^{\text{ \ \ }D}C_{DE}^{\text{ \ \ }F}C_{CF}^{\text{ \ \ }E}%
=C_{BC}^{\text{ \ \ }D}C_{AE}^{\text{ \ \ }F}C_{DF}^{\text{ \ \ }E}.
\end{equation}
This means that the invariance of Killing-Cartan product $\left(  X,X\right)
_{S-\exp}$ leads to the following condition condition $S$ for semigroup:
\begin{equation}
K_{\alpha\beta}^{\delta}K_{\delta\epsilon}^{\varepsilon}K_{\gamma\varepsilon
}^{\epsilon}=K_{\beta\gamma}^{\delta}K_{\alpha\epsilon}^{\varepsilon}%
K_{\delta\varepsilon}^{\epsilon}. \label{one'}%
\end{equation}

On the other hand, in Ref. \cite{irs} was shown that the $n$-selectors
2-selectors $K_{\alpha\beta}^{\gamma}$ of the $S$ semigroup satisfy the properties,%

\begin{equation}
K_{\alpha_{1}....\alpha_{n-1}}^{\sigma}K_{\sigma\alpha_{n}}^{\rho}%
=K_{\alpha_{1}\sigma}^{\rho}K_{\alpha_{2}....\alpha_{n}}^{\sigma}%
=K_{\alpha_{1}\alpha_{2}....\alpha_{n}}^{\rho},
\end{equation}%
\begin{equation}
K_{\alpha\beta}^{\delta}K_{\delta\epsilon}^{\varepsilon}=K_{\alpha\delta
}^{\varepsilon}K_{\beta\epsilon}^{\delta}=K_{\alpha\beta\epsilon}%
^{\varepsilon},
\end{equation}%
\[
K_{\alpha\beta}^{\delta}K_{\delta\epsilon}^{\varepsilon}=K_{\alpha\delta
}^{\varepsilon}K_{\beta\epsilon}^{\delta}=K_{\alpha\delta}^{\varepsilon
}K_{\epsilon\beta}^{\delta}=K_{\alpha\epsilon\beta}^{\varepsilon},
\]%
\[
K_{\beta\alpha\epsilon}^{\varepsilon}=K_{\alpha\beta\epsilon}^{\varepsilon
}=K_{\alpha\epsilon\beta}^{\varepsilon}.
\]

Using the above properties we find%
\begin{equation}
K_{\alpha\beta}^{\delta}K_{\delta\epsilon}^{\varepsilon}K_{\gamma\varepsilon
}^{\epsilon}=K_{\alpha\beta\epsilon}^{\varepsilon}K_{\gamma\varepsilon
}^{\epsilon}=K_{\alpha\beta\epsilon}^{\varepsilon}K_{\varepsilon\gamma
}^{\epsilon}=K_{\alpha\beta\varepsilon\gamma}^{\varepsilon}=K_{\alpha
\beta\gamma\varepsilon}^{\varepsilon}%
\end{equation}%
\begin{equation}
K_{\beta\gamma}^{\delta}K_{\alpha\epsilon}^{\varepsilon}K_{\delta\varepsilon
}^{\epsilon}=K_{\beta\gamma}^{\delta}K_{\alpha\delta\varepsilon}^{\varepsilon
}=K_{\beta\gamma}^{\delta}K_{\delta\alpha\varepsilon}^{\varepsilon}%
=K_{\beta\gamma\alpha\varepsilon}^{\varepsilon}=K_{\beta\alpha\gamma
\varepsilon}^{\varepsilon}=K_{\alpha\beta\gamma\varepsilon}^{\varepsilon}%
\end{equation}
and therefore%
\begin{equation}
K_{\alpha\beta}^{\delta}K_{\delta\epsilon}^{\varepsilon}K_{\gamma\varepsilon
}^{\epsilon}=K_{\beta\gamma}^{\delta}K_{\alpha\epsilon}^{\varepsilon}%
K_{\delta\varepsilon}^{\epsilon}. \label{two'}%
\end{equation}

Comparing the result (\ref{one'}) with (\ref{two'}), we conclude that the
inner product of Killing-Cartan $\left(  X,X\right)  _{S-\exp}$ is invariant
under linear transformations generated by $S\otimes\mathcal{G}.$

\end{document}